\documentclass[12pt]{article}
\usepackage{amsmath,amssymb,amsfonts}
\usepackage{mathrsfs}
\usepackage{bm}
\usepackage[english]{babel}
\usepackage{csquotes}
\usepackage{graphicx}
\usepackage{float}
\usepackage{booktabs}
\usepackage{multirow}
\usepackage{colortbl}
\usepackage{caption}
\usepackage{subcaption}
\usepackage{xcolor}
\usepackage[dvipsnames]{xcolor}
\usepackage{ulem}
\usepackage[DIV=13]{typearea}
\usepackage{ragged2e}
\usepackage{cite}
\usepackage{hyperref}
\hypersetup{colorlinks=true,urlcolor=magenta,anchorcolor=blue,citecolor=blue,filecolor=blue,linkcolor=magenta,menucolor=blue, linktocpage=true,pdfproducer=medialab}
\usepackage{cleveref}

\newcommand{\email}[1]{\href{mailto:#1}{\tt #1}}

\numberwithin{equation}{section}

\newcommand\subsetsim{\mathrel{%
  \ooalign{\raise0.2ex\hbox{$\subset$}\cr\hidewidth\raise-0.8ex\hbox{\scalebox{0.9}{$\sim$}}\hidewidth\cr}}}

\newcommand{\blue}[1]{\color{blue} #1 \color{black}}
\newcommand{\magenta}[1]{\color{magenta} #1 \color{black}}
\newcommand{\be}{\begin{equation}}
\newcommand{\ee}{\end{equation}}
\newcommand{\ba} {\begin{equation}\begin{aligned}}
\newcommand{\ea} {\end{aligned}\end{equation}}

\newcommand{\sL}{\mathscr{L}}

\newcommand{\cO}{\mathcal{O}}

\newcommand{\hc}{\text{h.c.}}
\newcommand{\nn}{\nonumber}
\newcommand{\vev}[1]{\langle #1\rangle}
\newcommand{\ov}[1]{\overline{#1}}

\def\Re{{\texttt{Re}}}

\newcommand{\TeV}{\ \text{TeV}}
\newcommand{\GeV}{\ \text{GeV}}

\newcommand{\alpaca}{{\fontfamily{cmss}\selectfont ALP-aca}}

\def\meg{{\mu\to e\gamma}}

\def\mNeN{{\mu^-N\to e^-N}}
\def\teg{{\tau\to e\gamma}}
\def\tmg{{\tau\to \mu\gamma}}
\def\tte{{\tau\to 3e}}
\def\ttm{{\tau\to 3\mu}}

\renewcommand*{\thefootnote}{\fnsymbol{footnote}}

\begin{document}

\begin{titlepage}

\vspace*{-1cm}
\flushleft{\magenta{IFT-UAM/CSIC-26-21}} 
\\[1cm]
\vskip 1cm

\begin{center}
\blue{\boldmath\bf \Large Crossing into the $m_a > f_a$ Region}
\vskip .3cm
\blue{\boldmath\bf \Large for Leptophilic ALPs}
\vskip .3cm
\end{center}

\vskip 0.5cm

\begin{center}
{\large\bf Marta F. Zamoro}~\footnote{\email{marta.zamoro@uam.es}},
{\large\bf \'Alvaro Lozano-Onrubia}~\footnote{\email{alvaro.lozano.onrubia@csic.es}},\\
\vspace{0.5cm}
{\large\bf Luca Merlo}~\footnote{\email{luca.merlo@uam.es}}, 
and
{\large\bf Samuel Rosende Herrero}~\footnote{\email{samuel.rosende@uam.es}}
\vskip .7cm
{\footnotesize
Departamento de F\'isica Te\'orica and Instituto de F\'isica Te\'orica UAM/CSIC,\\
Universidad Aut\'onoma de Madrid, Cantoblanco, 28049, Madrid, Spain
}
\end{center}

\vskip 2cm

\begin{abstract}
\justify
Axion-like particles (ALPs) are typically identified as pseudoscalars whose couplings are shift-symmetry invariant with the exception of their couplings to gauge bosons and their mass term. Additionally, the ALP mass $m_a$ is usually assumed to be (much) smaller than the ALP decay constant $f_a$. The latter condition is conservative, at best, and excludes part of the ALP parameter space that is presently viable. We revisit the criterion of the $m_a\ll f_a$ and perform an analysis focussing on leptophilic ALPs. In particular, we explore regions of the parameter space still uninvestigated, where $m_a>f_a$, thus providing a phenomenological study of the ALP-lepton couplings complementary to the existing literature. We point out that a leptophilic ALP may explain the $-3.8\sigma$ tension in the anomalous magnetic dipole moment of the electron for the Caesium determination in a large region of the $m_a\times f_a$ parameter space, testable in the near future through studies on $\mu\to e$ conversion in nuclei.
\end{abstract}

\end{titlepage}

\setcounter{footnote}{0}
\renewcommand*{\thefootnote}{\arabic{footnote}}

\pdfbookmark[1]{Table of Contents}{tableofcontents}
\tableofcontents


\section{Introduction}
\label{sec:Intro}

Despite its overwhelming success in describing fundamental particles and their interactions, the Standard Model of Particle Physics (SM) presents a number of striking shortcomings. One of the most prominent is the so called Strong CP problem, namely the absence of CP violating (CPV) sources in the strong sector of the SM, independent of the experimentally established CPV sources in the weak sector, i.e.\ those encountered in fermion mixing and embedded in the Cabibbo-Kobayashi-Maskawa (CKM) and Pontecorvo-Maki-Nakagawa-Sakata (PMNS) mixing matrices. One of the most accepted solutions to this problem is the QCD axion model~\cite{Peccei:1977hh,Weinberg:1977ma,
Wilczek:1977pj,Zhitnitsky:1980tq,Dine:1981rt,Kim:1979if,Shifman:1979if}. In short, the particle spectrum is extended beyond the SM (BSM) to include new scalar and/or fermionic fields, and the SM symmetry group is supplemented by a global $U(1)_\text{PQ}$ factor, the so-called Peccei-Quinn (PQ) symmetry, which by construction shall be anomalous with respect to QCD. Once the PQ symmetry is spontaneously broken, a new Nambu-Goldstone Boson (NGB) termed the axion arises. This NGB can absorb any additional CPV sources beyond the CKM and PMNS ones, representing a dynamical solution to the Strong CP problem. In all existing realisations of this solution the axion turns out to be a pseudo-NGB (pNGB) due to non-perturbative QCD effects that give rise to an axion mass $m_a$. This QCD axion mass satisfies a strict inverse proportionality relation with its characteristic scale $f_a$, which suppresses all its interactions:
\be
m_a f_a\approx m_\pi f_\pi\,,
\label{AxionMassRelation}
\ee
where $m_\pi$ and $f_\pi$ are the pion mass and its decay constant (for a formal derivation see for instance Ref.~\cite{GrillidiCortona:2015jxo}, also, recently in Ref.~\cite{Brandt:2026wir} the axion-pion mixing contribution to the photon coupling has been derived using lattice methods). 

The low-energy phenomenology of the different QCD axion models can be described in full generality in terms of an effective field theory (EFT). Following the pioneering article Ref.~\cite{Georgi:1986df}, the axion is described as a pNGB whose interactions respect a shift symmetry with the exception of its couplings to the gauge bosons, known as anomalous terms, and its mass term that follows the relation in Eq.~\eqref{AxionMassRelation}, in accordance with the Chiral Perturbation Theory ($\chi$PT) treatment. The coetaneous idea of a particle with properties closely resembling those of the QCD axion, with the crucial exception of the axion mass relation, had already surfaced in other contexts. Hence, the effective description of Ref.~\cite{Georgi:1986df} was assimilated elsewhere too. Thus emerged the axion-like-particle (ALP) to denominate any particle with interactions described by shift-symmetric couplings, anomalous terms, and a free mass that specifically does not need to satisfy the relation in Eq.~\eqref{AxionMassRelation}.

The literature presents a rich variety of ALPs: they show up in composite Higgs models~\cite{Merlo:2017sun,Brivio:2017sdm,
Alonso-Gonzalez:2018vpc,Alonso-Gonzalez:2020wst}, in supersymmetric constructions~\cite{Bellazzini:2017neg} and in the context of string theory \cite{Witten:1984dg,Choi:2006qj,
Svrcek:2006yi,Arvanitaki:2009fg,Cicoli:2012sz}, among others. ALPs have also been proposed as possible Dark Matter candidates~\cite{Gelmini:1984pe,Berezinsky:1993fm,Lattanzi:2007ux,
Bazzocchi:2008fh,Lattanzi:2013uza,Queiroz:2014yna} and, more generally, have been thought of as having an impact on cosmological observables~\cite{Ferreira:2018vjj,DEramo:2018vss,Escudero:2019gvw,Arias-Aragon:2020qtn,Arias-Aragon:2020qip, Arias-Aragon:2020shv,Ferreira:2020bpb,Escudero:2021rfi,Araki:2021xdk,DEramo:2021psx,DEramo:2021lgb, DEramo:2022nvb}. Furthermore, they have been considered in flavour model building~\cite{Davidson:1981zd,Wilczek:1982rv,Ema:2016ops, Calibbi:2016hwq,Arias-Aragon:2017eww,Arias-Aragon:2022ats,DiLuzio:2023ndz,Greljo:2024evt} and are known to emerge in mechanisms of neutrino mass generation, where they go by the name of Majorons~\cite{Chikashige:1980qk,Chikashige:1980ui,Gelmini:1980re,deGiorgi:2023tvn,Liang:2024vnd,Greljo:2025suh}. A recent and complete review that includes models with ALPs is Ref.~\cite{Albertus:2026fbe}.

In addition to the model building efforts, part of the community has extensively worked to improve and complete the ALP effective description: the literature ranges from new theoretical perspectives~\cite{Choi:1986zw,Salvio:2013iaa,
Brivio:2017ije,Alonso-Alvarez:2018irt,Gavela:2019wzg,Chala:2020wvs,Bonilla:2021ufe,Arias-Aragon:2022byr,
Arias-Aragon:2022iwl,DiLuzio:2023lmd,DiLuzio:2023cuk,deLima:2026bks} to more experimentally oriented analyses of potential ALP signals at both colliders~\cite{Jaeckel:2012yz,Mimasu:2014nea,
Jaeckel:2015jla,Alves:2016koo,Knapen:2016moh,Brivio:2017ije,Bauer:2017nlg,Mariotti:2017vtv,Bauer:2017ris,
Baldenegro:2018hng,Craig:2018kne,Bauer:2018uxu,Gavela:2019cmq,Haghighat:2020nuh,Wang:2021uyb,deGiorgi:2022oks,
Bonilla:2022pxu,Ghebretinsaea:2022djg,Vileta:2022jou,Calibbi:2022izs,Marcos:2024yfm,Arias-Aragon:2024gpm,Biekotter:2025fll,Ema:2025bww} and low-energy 
facilities~\cite{Izaguirre:2016dfi,Marciano:2016yhf,Merlo:2019anv,Aloni:2019ruo,Bauer:2019gfk,Cornella:2019uxs,Bauer:2020jbp,Calibbi:2020jvd,DiLuzio:2020oah,Bauer:2021mvw,Carmona:2021seb,Guerrera:2021yss,Gallo:2021ame,Bertholet:2021hjl,Cheng:2021kjg,Bonilla:2022qgm,Bonilla:2022vtn,deGiorgi:2022vup,Guerrera:2022ykl,Bonilla:2023dtf,Arias-Aragon:2023ehh,DiLuzio:2024jip,deGiorgi:2024str,Alda:2024cxn,Alda:2024xxa,Arias-Aragon:2024qji,Arias-Aragon:2024gdz,Bisht:2024hbs,Calibbi:2024rcm,MartinCamalich:2025srw,Alda:2025uwo,Alda:2025nsz,Ardu:2026vsr,Eberhart:2025lyu}. A very recent review on this topic can be found in Ref.~\cite{Arza:2026rsl}.

Contrary to the common understanding that a generic ALP cannot solve the Strong CP problem, recent results demonstrate~\cite{DiLuzio:2016sbl,DiLuzio:2017pfr,Gaillard:2018xgk,Hook:2019qoh,DiLuzio:2020wdo,DiLuzio:2020oah,DiLuzio:2021pxd,DiLuzio:2021gos,Gavela:2023tzu,Cox:2023dou,deGiorgi:2024elx} that the inverse proportionality between the mass $m_a$ and the scale $f_a$ traditionally touted for the QCD axion can be relaxed. It follows that the parameter space accommodating a consistent solution to the Strong CP problem exceeds that of the very constrained QCD axion band. Specifically, this allows one to consider ALPs as effective descriptions of axions that indeed solve the Strong CP problem yet whose mass is not described by Eq.~\eqref{AxionMassRelation}.

In accordance with the traditional QCD axion perspective, most of the ALP literature \textit{a priori} assumes the hierarchy of scales
\begin{equation}
\label{centralequation}
    m_a\ll f_a \, .
\end{equation}
This identification of the characteristic scale $f_a$ as an apparent cut-off scale appears naively reasonable on two fronts. First, ALP EFTs constitute a textbook instance of a bottom-up EFT approach wherein the connection to the actual EFT cut-off scale $\Lambda$ is not necessarily made manifest. This occurs because the approach embodied by ALP EFTs has $f_a$, and not the cut-off $\Lambda$, set an apparent scale of validity for the EFT (see Sec.~\ref{sec:DimAnalysis}). Although this may be viewed as a notational rather than a fundamentally physical issue, it is relevant to recall that the relation between $f_a$ and $\Lambda$ cannot, in general, be established if the EFT approach is truly ultraviolet (UV) agnostic.  Since any sensible ALP EFT will require $f_a \leq \Lambda$, the assumption in Eq.~\eqref{centralequation} ensures a physical mass well below the true scale of validity and thus the consistency of the EFT. Furthermore, Eq.~\eqref{centralequation} also happens to be an experimentally justified working hypothesis in contexts where the ALP is assumed light with respect to the Fermi scale. Indeed, the experimental bounds on the actual EFT cut-off scale $\Lambda$, resulting from both collider and flavour facilities, turn out to be in the TeV region for $\mathcal{O}(1)$ Wilson coefficients. As we shall later see, $f_a$ is not expected to veer off far below that scale unless an extremely strongly coupled UV theory is assumed.

However, caution is needed when blindly assuming the general relation Eq.~\eqref{centralequation} for two reasons. On the one hand, the scale $\Lambda$ can be much smaller than traditionally assumed. Indeed, if the Wilson coefficients happened to be substantially suppressed, $\Lambda$ would approach the scale of a traditionally light ALP mass. On the other hand, and all prejudice aside, an ALP may well be very massive when compared to other scales in the standard ALP EFT description. There is no fundamental reason why $m_a$ should be strictly constrained by the value of $f_a$ before the latter is set in relation to an EFT cut-off $\Lambda$. Poignantly, particle physics already showcases a striking example of an EFT featuring pNGBs with masses $m_{\pi}$ larger than their associated characteristic decay scale $f_{\pi}$, namely pions in $\chi$PT. Another example is the model proposed in Ref.~\cite{Murayama:2026ioh}, recently introduced as a solution to the Strong CP problem featuring a GeV-scale axion. The model is currently under scrutiny by the community~\cite{DiLuzio:2026poh} and appears to be excluded. Interestingly, the exclusion arises from its incompatibility with experimental data on pion physics, rather than from a theoretical inconsistency.

This observation alone calls for prudence, and, all things considered, a fundamental reappraisal of the working hypothesis $m_a\ll f_a$ prevalent in the literature appears necessary. 

In the next sections, {\it we claim that the direct comparison between $m_a$ and $f_a$ should be taken with a grain of salt and we exemplify the consequences of relaxing the $m_a\ll f_a$ condition in a simplified scenario.} The study focusses on a leptophilic ALP whose dominant phenomenology resides in charged lepton decays and in the magnetic dipole moment of the electron and the muon. While most of the studies of ALPs assume the parametric hierarchy $m_a\ll f_a$, being $m_a$ either very light or even in the GeV-range~\cite{Jaeckel:2012yz,Mimasu:2014nea,
Jaeckel:2015jla,Alves:2016koo,Knapen:2016moh,Brivio:2017ije,Bauer:2017nlg,Mariotti:2017vtv,Bauer:2017ris,
Baldenegro:2018hng,Craig:2018kne,Bauer:2018uxu,Gavela:2019cmq,Haghighat:2020nuh,Wang:2021uyb,deGiorgi:2022oks,
Bonilla:2022pxu,Ghebretinsaea:2022djg,Vileta:2022jou,Calibbi:2022izs,Marcos:2024yfm,Arias-Aragon:2024gpm,Biekotter:2025fll,Ema:2025bww,Izaguirre:2016dfi,Marciano:2016yhf,Merlo:2019anv,Aloni:2019ruo,Bauer:2019gfk,Cornella:2019uxs,Bauer:2020jbp,Calibbi:2020jvd,DiLuzio:2020oah,Bauer:2021mvw,Carmona:2021seb,Guerrera:2021yss,Gallo:2021ame,Bertholet:2021hjl,Cheng:2021kjg,Bonilla:2022qgm,Bonilla:2022vtn,deGiorgi:2022vup,Guerrera:2022ykl,Bonilla:2023dtf,Arias-Aragon:2023ehh,DiLuzio:2024jip,deGiorgi:2024str,Alda:2024cxn,Alda:2024xxa,Arias-Aragon:2024qji,Arias-Aragon:2024gdz,Bisht:2024hbs,Calibbi:2024rcm,MartinCamalich:2025srw,Alda:2025uwo,Alda:2025nsz,Ardu:2026vsr,Eberhart:2025lyu,Escribano:2020wua}, {\it we will focus on the $m_a> f_a$ region, that has hitherto received little attention in phenomenological analyses, thus providing results complementary to the existing investigations. As a part of this analysis, we point out the possibility to completely relax the present $-3.8\sigma$ tension between the Caesium determination of the anomalous magnetic dipole moment of the electron and its current experimental measurement within the context of a leptophilic ALP.} 

The structure of the paper, as described in the table of contents, includes in Sect.~\ref{sec:DimAnalysis} a review of  the relations between $m_a$, $f_a$ and $\Lambda$ within the Naive Dimensional Analysis description, followed in Sect.~\ref{sec:EFT} by a description of the EFT Lagrangian that represents our starting point for the analysis. In Sect.~\ref{sec:observables}, we list the physical processes that will be used in the numerical study, present the explicit dominant contributions to the expression of their observables, and identify specific benchmark textures for the ALP-lepton couplings. Sect.~\ref{sec:results} illustrates the results of our numerical analysis, first focussing on the different flavour textures and then on the possible explanation of the $(g-2)_e$ anomaly. The \alpaca\ code~\cite{Alda:2025nsz} is the main tool adopted for this analysis, in order to correctly account for the Renormalisation Group Equations (RGE) contributions. We then conclude in Sect.~\ref{sec:Conclusions}. Additional material on the RGE analysis can be found in App.~\ref{app:Running}, and a supplementary Mathematica file provides the complete expressions for the observables used, including a comparison with the available literature.

\section{Dimensional Analysis and Coupling Regime}
\label{sec:DimAnalysis}

In this section we discuss the $m_a\ll f_a$ relation that is commonly assumed in the ALP effective description. At first sight, comparing $m_a$ with $f_a$ seems legitimate as both quantities have the same mass dimension. However, they fundamentally embody two different and independent concepts: $m_a$ is a mass parameter, while $f_a$ is a decay constant that governs the ALP couplings. This comparison of $m_a$ {\it vs.} $f_a$ is actually specific of ALP EFTs and not of UV constructions: considering for example the traditional QCD axion model, the relation in Eq.~\eqref{AxionMassRelation} is well-defined as the same types of quantities appear on both sides of the equation, i.e.\ a mass and a decay constant. To clarify the fundamental relation between these conceptually different quantities, we need to dive deeper into the details of the EFT construction.

Independently of the specific high-energy dynamics that gives rise to the ALP, the low-energy description only deals with particles lighter than the EFT cut-off $\Lambda$ which represents the scale at which heavier degrees of freedom have been integrated out. In EFTs, $\Lambda$ often coincides with the mass of the heavier degrees of freedom~\footnote{If the heavy fields contribute to the effective operators only at loop level, then the cut-off $\Lambda$ should not be directly identified with their mass. In such cases $\Lambda$ rather parametrises the suppression scale of the induced operators, which can differ from the heavy mass by loop factors (e.g. $\Lambda \sim 16\pi^2 M$).}, meaning that any good EFT description cannot describe particles with masses larger than $\Lambda$. The same general principle also applies to ALP EFTs: a sensible EFT description requires an ALP mass below the cut-off. Thus, $m_a<\Lambda$ is a general consistency condition for the low-energy description to be meaningful. Since both quantities refer to mass parameters, the comparison of $m_a$ with $\Lambda$ is not only legitimate but also conceptually valid.  

We now take a step forward considering the pNGB nature of the ALP, that is, as already stated in the Introduction, treating it as a real, CP-odd scalar field, whose interactions present a shift-symmetry invariance, except for the anomalous couplings and its mass term.
This aspect allows us to harness the well-known framework of Naive Dimensional Analysis (NDA)~\cite{Manohar:1983md,Cohen:1997rt,Gavela:2016bzc} that encodes the usual $\Lambda$ counting rule in EFT combined with the $4\pi$ counting rule~\footnote{The $4\pi$ counting is
equivalent to $\hbar$ counting in the EFT, as also discussed previously in Refs.~\cite{Elias-Miro:2013mua,Espinosa,Gavela:2016bzc}.}, altogether providing a systematic way to normalize EFT operators. The $4\pi$ counting rule is very useful in the ALP description due to its close relation to the nature of NGBs, as is the case of $\chi$PT. We can write the NDA master formula as follows,
\be
\dfrac{\Lambda^4}{g_\ast^2}
\left[\dfrac{\partial}{\Lambda}\right]^{n_\partial}
\left[\dfrac{g_\ast \phi}{\Lambda}\right]^{n_\phi}
\left[\dfrac{g_\ast A}{\Lambda}\right]^{n_A}
\left[\dfrac{g_\ast \psi}{\Lambda^{3/2}}\right]^{n_\psi}
\label{NDAMasterFormula}
\ee
where $n_i$ is the number of $i$-element insertions in any effective operator, being $\phi$ any scalar field, $A$ any gauge boson, and $\psi$ any fermion.
In particular, $g_\ast$ is a measure of the strength of the underlying dynamics that gives rise to the low-energy operators: $g_\ast=4\pi$ represents the strongly interacting regime, while $g_\ast=1$ stands for the weakly interacting one, with the intermediate values interpolating between these two limits and characterising couplings of moderate strength.

In the case of $\chi$PT and focussing on the simpler case of pions, the Lagrangian is written in terms of $U\equiv \text{exp}\, 2i{\bm\pi}/f_\pi$, where $\bm\pi$ is the pion matrix and $f_\pi$ is the pion decay constant. Expanding $U$ in terms of $f_\pi$ implies that any meson insertion comes with a $1/f_\pi$ factor, i.e.\ an insertion rule given by $[{\bm\pi}/f_\pi]^{n_{\bm\pi}}$. We can thus compare with Eq.~\eqref{NDAMasterFormula} and identify a relation between $\Lambda$ and $f_\pi$,
\be
\Lambda=g_\ast f_\pi\,.
\label{Lambdafpi}
\ee
This has been already pointed out in the pioneering article by Georgi and Manohar~\cite{Manohar:1983md} and beautifully makes the point that the only meaningful scale in the game is the cut-off $\Lambda$ and not $f_\pi$. Focussing for instance on pion scattering, the derivative expansion characteristic of $\chi$PT is an expansion in powers of $\partial/\Lambda$ that is valid for small pion momenta up to $\Lambda=4\pi f_\pi\sim1\GeV$. If instead of $\Lambda$ we used $f_\pi$ as the cut-off of the theory and thus as the momentum expansion parameter, the fact that $m_\pi>f_\pi$ would result in the failure of $\chi$PT for pion scattering already at the threshold $p\sim m_\pi$. Evidently, this not the case.

We can easily generalise the previous discussion to the ALP scenario: although the characteristic scale suppressing the ALP couplings can be taken to be $f_a$, the scale that defines the validity of the EFT description is $\Lambda$. As in the case of the pions, there is no contradiction in the possibility of $m_a$ larger than $f_a$ as long as $m_a$ remains below $\Lambda$.
\\

Despite the above discussion, we note that the comparison between $m_a$ and $f_a$ can still provide useful information. If an ALP signal consistent with a mass larger than $f_a$ is detected, this suggests that the underlying dynamics may be strongly interacting. Likewise, if the signal points to $m_a\lesssim f_a$, then a weakly interacting UV shall be favoured as the origin of the ALP. We can exemplify this concept adopting the NDA master formula Eq.~\eqref{NDAMasterFormula} as applied to a simple model containing a complex scalar field $\phi(x)$ transforming under the PQ symmetry. The associated Lagrangian reads
\be
\sL\supset\partial_\mu \phi^\ast\, \partial^\mu \phi -\Big[-\Lambda^2\phi^\ast\phi+g_\ast^2\left(\phi^\ast\phi\right)^2\Big]\,,
\label{ScalarLagExample}
\ee
and when $g_\ast=4\pi$ the 1-loop corrections to the quartic couplings are as large as the tree-level parameter, thus describing a strongly interacting regime. 

At the minimum,
\begin{equation}
\phi(x) = \dfrac{\vev{\phi}+\rho(x)}{\sqrt{2}} e^{i\,a(x)/f_a}\,,\qquad
\vev{\phi}=\dfrac{\Lambda}{g_\ast}\,,
\end{equation}
with $\rho(x)$ the radial degree of freedom with mass $\Lambda$ and $a$ is the ALP.
Once we focus on the interactions describing the ALP, canonical kinetic terms imply a relation between $f_a$ and $\vev{\phi}$,  that is $f_a=\vev{\phi}$. This relation is equivalent to the one in Eq.~\eqref{Lambdafpi} valid in $\chi$PT,
\be
\Lambda=g_\ast\,f_a\,.
\ee
Let's now add a derivative coupling with a generic fermion $\psi$: still according to the NDA master formula, 
\be
\dfrac{g_\ast}{\Lambda}\partial_\mu\phi \,\ov{\psi}\psi= 
\dfrac{1}{f_a}\partial_\mu\phi \,\ov{\psi}\psi\Longrightarrow 
\dfrac{1}{f_a}\partial_\mu a\, \ov{\psi}\psi\,,
\label{DervALPCouplGeneric}
\ee
where in the last step we retained only the ALP degree of freedom. 

This minimal and simplified example leads to the idea that {\it the ALP may arise either from a weakly or a strongly interacting regime, but the form of the ALP interactions are insensitive to this aspect. As a consequence, analyses of ALP couplings are not able to tell anything about the $g_\ast$ factor of Eq.~\eqref{Lambdafpi} that instead can be investigated with direct or indirect searches of the radial mode $\rho$. In contrast, if the ALP mass is larger than $f_a$, this alone signals that the particle must arise from strong dynamics. It is important to stress, however, that the reverse implication fails: $m_a < f_a$ does not guarantee a weakly interacting ALP.}


\section{The Effective Description}
\label{sec:EFT}

Our starting point is the effective ALP Lagrangian in the so-called chirality-preserving or derivative basis, which at the cut-off scale $\Lambda$ can be written as:
\begin{align}
\sL_{\partial a}(\Lambda) ={}&\dfrac{1}{2}(\partial_\mu a)^2 - \dfrac{1}{2}m_a^2 a^2
+\dfrac{a}{f_a}\left(\dfrac{g'^2 c_{\widetilde B}}{16\pi^2} B_{\mu\nu}\widetilde{B}^{\mu\nu}
+\dfrac{g^2 c_{\widetilde W}}{16\pi^2} W^i_{\mu\nu}\widetilde{W}^{i,\mu\nu}
+\dfrac{g_s^2 c_{\widetilde G}}{16\pi^2}G^a_{\mu\nu}\widetilde{G}^{a,\mu\nu}\right)\nonumber\\
&+\dfrac{\partial^\mu a}{f_a}\left(\ov{q'_L}\gamma_\mu c'_{q}q'_L+\ov{u'_R}\gamma_\mu c'_{u}u'_R
+\ov{d'_R}\gamma_\mu c'_{d}d'_R+\ov{\ell'_L}\gamma_\mu c'_{\ell}\ell'_L+\ov{e'_R}\gamma_\mu c'_{e}e'_R\right)\,.
\label{eq:UVLag}
\end{align}
to first order of the expansion in $1/f_a$.
Here $g_s$, $g$ and $g'$ denote the strong, weak and hypercharge couplings respectively; $G^a_{\mu\nu}$, $W^i_{\mu\nu}$ 
and $B_{\mu\nu}$ are the $\mathrm{SU}(3)_C \times \mathrm{SU}(2)_L \times \mathrm{U}(1)_Y$ field strengths and $\widetilde{X}^{\mu\nu}\equiv 
\dfrac{1}{2} \epsilon^{\mu\nu\rho\sigma} {X}_{\rho\sigma}$, with $\epsilon^{0123}=1$, their associated dual fields. The left-handed (LH) and right-handed (RH) fermions $f'=\{q'_L,u'_R,d'_R,\ell'_L,e'_R\}$ are triplets in flavour space, 
while the corresponding ALP couplings, $c'_f=\{c'_{q},c'_{u},c'_{d},c'_{\ell},c'_{e}\}$, are general $3\times 3$ hermitian 
matrices defined in the fermion flavour basis. No ALP--Higgs operator has been included in the Lagrangian of Eq.~\eqref{eq:UVLag}, being 
redundant when ALP--fermion operators are included \cite{Georgi:1986df}.

Performing unitary transformations on the fermion fields, beyond affecting the SM Lagrangian, induces a change in the second line of Eq.~\eqref{eq:UVLag}. Denoting the transformations as $\psi'_L=V_\psi\psi_L$ and $\psi'_R=U_\psi\psi_R$, being $V_\psi$ and $U_\psi$ unitary $3\times3$ matrices in the flavour space, the ALP-fermion couplings conserve the same form but with new coefficients $c_\psi$:
\begin{equation}
\begin{gathered}
c_u\equiv U_u^\dagger c'_u U_u\,,\qquad c_d\equiv U_d^\dagger c'_d U_d\,,\qquad c_e\equiv U_e^\dagger c'_e U_e\,,\\
c_q\equiv V_q^\dagger c'_q V_q\,,\qquad c_\ell\equiv V_\ell^\dagger c'_\ell V_\ell\,.
\end{gathered}
\label{cpsicppsigeneral}
\end{equation}

This will be useful when discussing the change to any other flavour basis or the fermion mass basis. 

It is possible to write the Lagrangian in Eq.~\eqref{eq:UVLag} in the so-called chirality-flipping or Yukawa-like basis, that has some advantages in the phenomenological study. By performing fermion chiral transformations, we can obtain
\begin{align}
\sL_a(\Lambda) = &\dfrac{1}{2}(\partial_\mu a)^2 -
\dfrac{1}{2}m_a^2 a^2 +
\dfrac{a}{f_a} 
\left(
\dfrac{g^{\prime 2} \hat{c}_{\widetilde B}}{16\pi^2} 
B_{\mu\nu}\widetilde{B}^{\mu\nu} +
\dfrac{g^2 \hat{c}_{\widetilde W}}{16\pi^2} 
W^i_{\mu\nu}\widetilde{W}^{i,\mu\nu} +
\dfrac{g_s^2 \hat{c}_{\widetilde G}}{16\pi^2}
G^a_{\mu\nu}\widetilde{G}^{a,\mu\nu}
\right) +\nn\\ 
&-\dfrac{a}{f_a} \left( \ov{q'_L}H \widehat{Y}'_d d'_R + 
\ov{q'_L} \widetilde{H} \widehat{Y}'_u u'_R + \ov{\ell'_L}H\widehat{Y}'_ee'_R + \hc
\right)\,.
\label{eq:UVLagRedoneNew}
\end{align}

In the previous expression, the ALP-fermion couplings are proportional to
\be
\widehat{Y}'_d = i \left( Y'_d c'_d - c'_q Y'_d \right)\,,\qquad
\widehat{Y}'_u = i \left( Y'_u c'_u - c'_q Y'_u\right)\,,\qquad
\widehat{Y}'_e = i \left( Y'_e c'_e - c'_{\ell} Y'_e \right)\,,
\ee
where $Y'_\psi$ are the Yukawa coupling matrices in the original flavour basis. On the other hand, the anomalous terms receive contributions from the fermion chiral transformations according to the Fujikawa method~\cite{Fujikawa:1979ay},
\be
\hat{c}_{\widetilde B} = c_{\widetilde B} + K_B\,,\qquad
\hat{c}_{\widetilde W} = c_{\widetilde W} + K_W\,,\qquad
\hat{c}_{\widetilde G} = c_{\widetilde G} +K_G\,,
\ee
where the $K_X$ are defined as
\be
\begin{gathered}
K_B \equiv  \textrm{Tr} \Bigg[ N_c\Big(N_L \mathcal{Y}_q^2c'_q-\mathcal{Y}_u^2 c'_u -\mathcal{Y}_d^2c'_d\Big) + N_L \mathcal{Y}_\ell^2 c'_{\ell} -\mathcal{Y}_e^2 c'_e\Bigg]\\
K_W \equiv  T_F \textrm{Tr} \Big[ N_c c'_q + c'_{\ell} \Big]\,,\qquad\qquad
K_G\equiv T_F \textrm{Tr} \Big[ N_L c'_q-c'_u - c'_d\Big]\,,
\end{gathered}
\label{Ks}
\ee
with $T_F = \dfrac{1}{2}$, $N_L=2$, $N_c=3$ and the hypercharges
\begin{align}
    \mathcal{Y}_q = \dfrac{1}{6}\,,\qquad
    \mathcal{Y}_u = \dfrac{2}{3}\,,\qquad
    \mathcal{Y}_d = -\dfrac{1}{3}\,,\qquad
    \mathcal{Y}_{\ell} = -\dfrac{1}{2}\,,\qquad
    \mathcal{Y}_e = -1 \, .
\end{align}

The Lagrangians in Eqs.~\eqref{eq:UVLag} and \eqref{eq:UVLagRedoneNew} are defined at the cut-off scale $\Lambda$. At lower scales, quantum corrections generically modify the ALP-fermion couplings. The Renormalisation Group Equations (RGE)~\cite{MartinCamalich:2020dfe,Bauer:2020jbp,Bonilla:2021ufe,DasBakshi:2023lca,Bresciani:2024shu} state that the Wilson coefficients $c_{\widetilde{B}}$, $c_{\widetilde{W}}$ and $c_{\widetilde{G}}$ of the anomalous terms do not receive RGE corrections at one loop, while the Wilson coefficients $c'_\psi$ do receive corrections at this order. Our main focus in this paper is to explore regions of the parameter space where $m_a > f_a$, keeping a conservative approach to the hierarchy between the PQ and the EW symmetry breaking, $\Lambda\gg v_\text{EW}$. For consistency, the condition $f_a>v_\text{EW}/4\pi$ must be respected, where $v_\text{EW}=246\GeV$ is the vacuum expectation value of the Higgs. 

At the end, the RGE running from the cut-off scale $\Lambda$ down to $m_a$ is taken into consideration, using the \alpaca~\cite{Alda:2025nsz} code to implement this part of the analysis. The relevant Lagrangian to match with depends on whether $m_a$ is larger or smaller than $v_\text{EW}$. In the former case, $m_a>v_\text{EW}$, the leading contributions to the fermionic Lagrangian read
\begin{align}
\sL^{>v}_{\partial a}(m_a) \supset{}&\dfrac{\partial^\mu a}{f_a}\left(\ov{q_L}\gamma_\mu c^L_q q_L+\ov{u_R}\gamma_\mu c^R_u u_R
+\ov{d_R}\gamma_\mu c^R_d d_R+\ov{\ell_L}\gamma_\mu c_\ell^L \ell_L+\ov{e_R}\gamma_\mu c_e^R e_R\right)\,.
\label{eq:UVLagma}
\end{align}
where $c^{L,R}$ are the running Wilson coefficients defined at the scale $m_a$. If, instead, $m_a<v_\text{EW}$ then the members of the $SU(2)_L$-doublets should be treated differently and the coefficients undergo a different running under the RGE. For example, focussing for simplicity on the lepton sector, the Lagrangian would contain three different terms,
\begin{equation}
\sL^{<v}_{\partial a}(m_a)\supset \dfrac{\partial^\mu a}{f_a}\left(\ov{e_L}\gamma_\mu\,c^L\,e_L+\ov{e_R}\,\gamma_\mu\,c^R e_R+\ov{\nu_L}\gamma_\mu\,c^\nu\,\nu_L\right)\,,
\end{equation}
where both $c^L$ and $c^\nu$ derive from $c_\ell^L$ and differ only for the RGE contributions.

For the purposes of this work, we restrict ourselves to leptonic observables, as they offer a cleaner and more precise ground compared to hadronic ones. {\it We will focus on leptophilic ALPs, i.e.\ ALPs that couple predominantly to leptons, with negligible couplings to quarks.} Nevertheless, running the leptonic couplings from a high scale $\Lambda$ down to the scale of the ALP mass $m_a$ will generate couplings to quarks, especially to the top ($t$) and bottom ($b$) flavours. Therefore, it is important to gauge the impact this could have on hadronic observables. For this purpose, we first perform the RGE running from the scale $\Lambda$ down to the ALP mass $m_a$, where the ALP is integrated out, and we match to the relevant effective Lagrangian at that scale, constructed only with SM fields. In the case of $m_a>v_{EW}$, the sensible description is the SMEFT~\cite{Buchmuller:1985jz,Grzadkowski:2010es} Lagrangian. 
We integrate out the ALP in the chirality-flipping basis of Eq.~\eqref{eq:UVLagRedoneNew} to simplify the procedure and, matching onto the SMEFT Warsaw basis~\cite{Grzadkowski:2010es} and applying Fiertz redefinitions, we find that the following $d=6$ operators receive contributions:
\be
\sL^{4\psi}_a=
~\mathcal{C}_{le}\mathcal{Q}_{le}+
\mathcal{C}_{qu}^{(1)}\mathcal{Q}_{qu}^{(1)}+
\mathcal{C}_{qd}^{(1)}\mathcal{Q}_{qd}^{(1)}
+\mathcal{C}_{ledq}\mathcal{Q}_{ledq}+\mathcal{C}_{quqd}^{(1)}\mathcal{Q}_{quqd}^{(1)}
+\mathcal{C}_{quqd}^{(8)}\mathcal{Q}_{quqd}^{(8)}+\mathcal{C}_{lequ}^{(1)}\mathcal{Q}_{lequ}^{(1)}
\label{EffectiveLag4FermionOperators}
\ee
where
\be
\begin{aligned}
\label{eq:wilson_coefficient}
\mathcal{C}_{le}=&
-\dfrac{v^2}{4 m_a^2 f_a^2}(
\widehat{Y}'_e)^{pt}(\widehat{Y}^{\prime\dag}_e)^{sr} 
\qquad\qquad
&
\mathcal{Q}_{le}=&
\left(\ov{e^{\prime p}_L} \gamma_\mu e_{L}^{\prime r}\right)
\left(\ov{e^{\prime s}_R}\gamma^\mu e_R^{\prime t}\right)\\
\mathcal{C}_{qu}^{(1)}=&
-\dfrac{v^2}{12 m_a^2 f_a^2}(\widehat{Y}'_u)^{pt}(\widehat{Y}^{\prime\dag}_u)^{sr}
\qquad\qquad
&
\mathcal{Q}_{qu}^{(1)}=&
\left(\ov{q^{\prime p}_L} \gamma_\mu q_L^{\prime r}\right)
\left(\ov{u^{\prime s}_R}\gamma^\mu u_R^{\prime t}\right)\\
\mathcal{C}_{qd}^{(1)}=&
-\dfrac{v^2}{12 m_a^2 f_a^2}(\widehat{Y}'_d)^{pt}(\widehat{Y}^{\prime\dag}_d)^{sr}
\qquad\qquad
&
\mathcal{Q}_{qd}^{(1)}=&
\left(\ov{q^{\prime p}_L} \gamma_\mu q_L^{\prime r}\right)
\left(\ov{d^{\prime s}_R}\gamma^\mu d_R^{\prime t}\right)\\
\mathcal{C}_{ledq}=&
-\dfrac{v^2}{2 m_a^2 f_a^2}(\widehat{Y}'_e)^{pr}(\widehat{Y}^{\prime\dag}_d)^{st} 
\qquad\qquad
&
\mathcal{Q}_{ledq}=&
\left(\ov{\ell^{\prime p}_L} e_R^{\prime r}\right) 
\left(\ov{d^{\prime s}_R}q^{\prime t}_L \right)\\
\mathcal{C}_{quqd}^{(1)}=&
\dfrac{v^2}{12 m_a^2 f_a^2}(\widehat{Y}'_u)^{sr} (\widehat{Y}'_d)^{pt}
\qquad\qquad
&
\mathcal{Q}_{quqd}^{(1)}=&
\left(\ov{q^{\prime p}_L} u_R^{\prime r}\right)\varepsilon_{jk} \left(\ov{q^{\prime s}_L} d_R^{\prime t}\right)\\
\mathcal{C}_{quqd}^{(8)}=&
\dfrac{v^2}{2 m_a^2 f_a^2}(\widehat{Y}'_u)^{sr} (\widehat{Y}'_d)^{pt}
\qquad\qquad
&
\mathcal{Q}_{quqd}^{(8)}=&
\left(\ov{q^{\prime p}_L} T^A u_R^{\prime r}\right)\varepsilon_{jk} \left(\ov{q^{\prime s}_L} T^A d_R^{\prime t}\right)\\
\mathcal{C}_{lequ}^{(1)}=&
\dfrac{v^2}{2 m_a^2 f_a^2}(\widehat{Y}'_e)^{pr}(\widehat{Y}'_u)^{st} 
\qquad\qquad
&
\mathcal{Q}_{lequ}^{(1)}=&
\left(\ov{\ell^{\prime p}_L} e_R^{\prime r}\right)\varepsilon_{jk} \left(\ov{q^{\prime s}_L} u_R^{\prime t}\right)\,,   
\end{aligned}
\addtocounter{equation}{-1}
\ee

where $T^A$ are the $SU(3)_c$ generators and $\varepsilon_{jk}$ is the 2-dimensional Levi-Civita tensor. We made explicit the flavour contractions in the expressions of the Wilson coefficients and of the operators. To the best of our knowledge, this has not been presented elsewhere in the literature, where the general treatment involves $m_a\ll v_\text{EW}$.

Observing the Wilson coefficients listed above, we can notice that most of them are strongly suppressed. On the one hand, there are suppressing factors of $v^2/f_a^2$ or $v^2/m_a^2$, depending on whether $f_a$ or $m_a$ constitutes the larger scale. On the other hand, the $\widehat{Y}'$ terms contain, by definition, the fermion Yukawas and the Lagrangian coefficients $c'_\psi$ of the ALP-fermion couplings. In particular, as already stated, we will focus on leptophilic ALPs and therefore the $c'_q$, $c'_u$ and $c'_d$ are vanishing at the scale $\Lambda$. Moreover, an explicit computation including the RGE effects reveals that the largest entries of the quark $\widehat{Y}'$ turn out to be of $\cO(10^{-4})$ and associated to the $t$ and 
$b$ flavours. We will enter more into details in the next section, after having defined the ALP-lepton coupling textures.  

On the whole, all Wilson coefficients $\mathcal{C}$ except for $\mathcal{C}_{le}$ are extremely suppressed and the corresponding operators turn out to have a negligible effect in our study. On the contrary, the operator $\mathcal{Q}_{le}$ may have a relevant role in the numerical analysis, depending on the specific ALP-lepton coupling texture considered. In the next section, we will indeed provide the explicit expressions for various charged lepton decays that can be described by this operator.

Regarding the ALP couplings to gauge bosons, and especially the coupling to photons, we remark that they play a fundamental role due to their effect on a large variety of observables, and will therefore be discussed in more detail in the next section. As mentioned before, the coefficients $c_{\widetilde{B}}$ and $c_{\widetilde{W}}$ do not receive any correction through the RGE running at one-loop (1L) level. Consistently with our hypothesis of leptophilic ALPs, where only the coefficients $c'_\ell$ and $c'_e$ in Eq.~\eqref{eq:UVLag} are non-vanishing, one may then conclude that no interactions with photons are then generated. However, this is not the case even at scales larger than $\Lambda_\text{QCD}$, the QCD confinement scale, where contributions are indeed expected due to the non-perturbative effects described within the Chiral Perturbation approach. The explicit computation of the $a\gamma\gamma$ coupling at 1L, described by the classic triangle Feynman diagram with internal charged leptons, is non-vanishing (see Ref.~\cite{Alda:2025uwo} for a complete and updated discussion on this aspect). In the end, the effective description of the ALP-photon couplings at any scale $m_a>m_\tau$ can be approximately written as
\begin{equation}
\sL^{<v}_{\partial a}(m_a)\supset\dfrac{a}{f_a}\dfrac{\alpha_\text{em}}{4\pi}\,c^\text{eff}_{a\gamma\gamma}F_{\mu\nu}\widetilde{F}^{\mu\nu}
\end{equation}
with
\begin{equation}
c^\text{eff}_{a\gamma\gamma}=-\dfrac{1}{2}\sum_{i}(c^L_{ii}-c^R_{ii})\,,
\label{ALPPhotonCouplingEFF}
\end{equation}
for the case under consideration of a leptophilic ALP. It will also be useful to specify the effective coupling of the ALP with one photon and one $Z$, as it enters in several penguin diagrams:
\begin{equation}
\sL^{<v}_{\partial a}(m_a)\supset\dfrac{a}{f_a}\dfrac{\alpha_\text{em}}{2\pi s_w c_w}\,c^\text{eff}_{a\gamma Z}F_{\mu\nu}\widetilde{Z}^{\mu\nu}
\end{equation}
with
\begin{equation}
c^\text{eff}_{a\gamma Z}=-9.4\times 10^{-3}\sum_{i} (c^L_{ii}-c^R_{ii})\,,
\end{equation}
for all the ALP mass values considered in our analysis, where $s_w$ and $c_w$ stand for the sine and cosine of the Weinberg angle, $\sin^2\theta_w=0.2312$~\cite{ParticleDataGroup:2024cfk}.\\

Before concluding this section, we comment on shift-symmetry-breaking terms and their possible effects. Let us again focus on the minimal scenario illustrated in the previous section, whose Lagrangian is given by Eq.~\eqref{ScalarLagExample}. By definition, any shift-symmetry-breaking dynamics can only operate below the scale at which the PQ symmetry is spontaneously broken, namely $\Lambda$. Naively, this is traditionally treated by introducing a “small” explicit symmetry breaking parameter, to which any such contribution is proportional. The advantage of this approach is that, once this parameter is set to zero, no explicit symmetry breaking effect survives. However, this does not imply that the resulting contributions to shift-symmetry-breaking observables must be of the same order as the small parameter itself. The pion mass again provides a useful example. Even neglecting electromagnetic interactions, the pion mass is much larger than the quark masses that explicitly break chiral symmetry. In this case, the dominant contribution to the pion mass arises from chromomagnetic interactions, which would ultimately vanish in the limit of massless quarks.

Generalising this discussion, the size of explicit shift-symmetry-breaking effects is in fact strongly model-dependent, and the ALP mass term does not necessarily play any special role beyond constituting the lowest dimensional operator that explicitly breaks the shift symmetry. Therefore, in our high-energy Lagrangian we shall not consider any other shift-symmetry-breaking terms beyond the anomalous couplings and the ALP mass term; without any specific underlying rationale, the opposite ansatz would have us dealing with a generic CP-odd scalar and not with an ALP.

\section{Observables}
\label{sec:observables}

Having established the theoretical discussion, we proceed with a data-driven approach. We will focus on leptonic processes to study the viability of the parameter space within the range $m_a> f_a$, although we will also integrate the complementary window $m_a < f_a$. As already stated, these observables offer a cleaner and more precise ground compared to hadronic ones. Although the analysis focusses on specific leptophilic ALP-coupling textures, it allows us to firmly establish that the $m_a>f_a$ parameter space is viable and interesting. Last but not least, this approach is self-consistent without the need of any {\it ad hoc} hypothesis, besides requiring a leptophilic ALP. In this case, the hadron phenomenology is negligible because the ALP-quark and ALP-gluon couplings are extremely suppressed.

The selected processes are 
\begin{itemize}
    \item The anomalous magnetic dipole moments of the electron and muon, $(g-2)_e$ and $(g-2)_\mu$.
    \item Charged lepton flavour-violating (cLFV) processes: radiative charged lepton decays, $\mu$--$e$ conversion in nuclei, and $\tau$ decays into lighter charged leptons.
\end{itemize}
They have a high experimental precision, especially when the electron and the muon are involved. A summary table with the SM predictions, current experimental values and future prospects is given in Tab.~\ref{tab:summary}. A comment is in order: when dealing with the $\mu$--$e$ conversion in nuclei, we consider the best present bound from SINDRUM II with gold, and the future prospects with aluminium from COMET. 

For the anomalous magnetic dipole moment of the electron we need to distinguish between the two determinations, involving the measurement of $\alpha_\textrm{em}$ with Caesium~\cite{Parker:2018vye} and with Rubidium~\cite{Morel:2020dww}, that lead to two different tensions with the current experimental measure: 
\be
\begin{aligned}
\delta a^\text{Cs}_e&=(-9.9\pm2.6)\times 10^{-13}\qquad (-3.8\sigma),\\
\delta a^\text{Rb}_e &=(3.7\pm1.5)\times 10^{-13}\hspace{10mm}(+2.4\sigma),
\end{aligned}
\ee
where $\delta a^\textit{X}_e\equiv a_e^\text{Exp}-a^\textit{X}_e$.

\begin{table}[ht!]
\hspace{-.5cm}
\begin{tabular}{ |c||c|c|  }
\hline
\parbox[c]{3cm}{\centering \textbf{Observable}} & 
\parbox[c]{4.5cm}{\centering \textbf{SM prediction $+\,\,m_\nu$}} &
\parbox[c]{6cm}{\centering \textbf{Experimental limits}} \\
\hline
& & \\
\multirow[c]{2}{*}{$(g-2)_e/2$} 
& \multirow[c]{2}{*}{\begin{tabular}{@{}c@{}} 
\text{Cs}:~0.00115965218161(23) \cite{Parker:2018vye} \\ 
\text{Rb}:~0.001159652180252(95) \cite{Morel:2020dww} 
\end{tabular}}
& 0.00115965218062(12) \cite{ParticleDataGroup:2024cfk} \\
& & \\[10pt]
$(g-2)_\mu/2$
&~0.00116592033(62) \cite{Aliberti:2025beg}
& 0.00116592059(22) \cite{Muong-2:2024hpx} \\
& & \\
\multirow[c]{2}{*}{$\mathcal{BR}(\mu \rightarrow e\gamma)$} 
& \multirow[c]{2}{*}{$\lesssim \mathcal{O}(10^{-55})$ \cite{Petcov:1977,cheng2000gauge,calibbi2018charged}} & $< 1.5\cdot 10^{-13}$ \cite{MEGII:2025gzr} \\
& & ($< 6 \cdot 10^{-14}$ \cite{MEGII:2023ltw,MEGII:2025gzr}) \\
& & \\
\multirow[c]{2}{*}{$\mathcal{BR}(\mu \rightarrow 3e)$} 
& \multirow[c]{2}{*}{$\lesssim \mathcal{O}(10^{-55})$ \cite{Hernandez-Tome:2018fbq,Petcov:1977}}
& $< 1.0\cdot 10^{-12}$ \cite{SINDRUM:1987nra} \\
& & (\text{Mu3e:}~$<5\cdot10^{-16}$ \cite{Redigolo:2024ztw}) \\
& & \\
\multirow[c]{3}{*}{$\mathcal{BR}(\mu + N \rightarrow e+N)$} 
& \multirow[c]{3}{*}{$\lesssim\mathcal{O}(10^{-54})$ \cite{Ilakovac:1994kj}}
& $\mathcal{BR}_{\text{Au}}  < 7 \cdot 10^{-13}$ \cite{SINDRUMII:2006dvw} \\
& & (\text{Mu2e:}~$\mathcal{BR}_{\text{Al}} < 2.87 \cdot 10^{-17}$ \cite{Mu2e:2014fns}) \\
& & (\text{COMET:}~$\mathcal{BR}_{\text{Al}} < 2 \cdot 10^{-17}$ \cite{Moritsu:2022lem}) \\
& & \\
$\mathcal{BR}(\tau\to e\gamma)$
&$\lesssim \mathcal{O}(10^{-49})$ \cite{cheng2000gauge}
& $<3.3\cdot 10^{-8}$ \cite{ParticleDataGroup:2024cfk} \\
& & \\
$\mathcal{BR}(\tau\to \mu\gamma)$
&$\lesssim \mathcal{O}(10^{-49})$ \cite{cheng2000gauge}
& $<4.2\cdot 10^{-8}$ \cite{ParticleDataGroup:2024cfk} \\
& & \\
$\mathcal{BR}(\tau\to 3e)$
&$\lesssim \mathcal{O}(10^{-56})$ \cite{Hernandez-Tome:2018fbq,Petcov:1977}
& $<2.7\cdot 10^{-8}$ \cite{ParticleDataGroup:2024cfk} \\
& & \\
$\mathcal{BR}(\tau\to 3\mu)$
& $\lesssim \mathcal{O}(10^{-55})$ \cite{Hernandez-Tome:2018fbq,Petcov:1977}
& $<1.9\cdot 10^{-8}$ \cite{ParticleDataGroup:2024cfk} \\
& & \\
$\mathcal{BR}(\tau^{-}\to e^{-}\mu^{+}\mu^{-})$
& $\lesssim \mathcal{O}(10^{-56})$ \cite{Hernandez-Tome:2018fbq,Petcov:1977}
& $<2.7\cdot 10^{-8}$ \cite{ParticleDataGroup:2024cfk} \\
& & \\
$\mathcal{BR}(\tau^{-}\to \mu^{-}e^{+}e^{-})$
& $\lesssim \mathcal{O}(10^{-55})$ \cite{Hernandez-Tome:2018fbq,Petcov:1977}
& $<1.8\cdot 10^{-8}$ \cite{ParticleDataGroup:2024cfk} \\
& & \\
\hline
\end{tabular}
\caption{\it Summary of leptonic observables used in our analysis, as well as their predictions within the SM context with massive active neutrinos and current experimental limits and the future prospects (in parentheses).}
\label{tab:summary}
\end{table}

Remarkably, the Caesium determination shows an anomaly that, if consolidated, would call for a New Physics contribution to be explained. We will discuss this possibility in the next section. 

For the $(g-2)_\mu$ the situation changed considerably in recent years. Indeed, the lattice-QCD calculation precision has increased and allows for a new precise determination of the hadron vacuum polarization contribution that results in a major upward shift of the total SM prediction. Compared against the current experimental determination, this implies that there is no tension between the SM and experiment at the current level of precision. Therefore, we will treat it at the same level as  the other constraints, and not as an anomaly.\\

Besides the above list of lepton flavour conserving and violating observables, one may further consider cosmological and astrophysical bounds, as well as collider constraints. Despite forgoing a dedicated analysis in this work, we provide solid arguments to justify not including both types of observables at present, whilst acknowledging that foreseeable improvements in precision of collider probes represent a promising venue for possible signals in the future.

Focussing first on astrophysical searches (for a review see, for instance, Part III of Ref.~\cite{Arza:2026rsl}) that include, in particular, studies on the ALP production mechanism in the Sun, red giants, neutron stars and supernovae,  very strong constraints can be derived from energy-loss arguments for ALP masses smaller than $m_a\sim 0.1\GeV$. Our study applies only to masses larger than this quantity and therefore we can avoid discussing these observables.  

On the other hand, for ALP masses in the GeV range or above, cosmological bounds are typically relevant only for extremely small ALP-photon couplings (for a review see, for instance, Part II of Ref.~\cite{Arza:2026rsl}). For larger couplings the ALP lifetime becomes very short, so that the particle decays well before epochs such as Big Bang Nucleosynthesis. In addition, these cosmological limits are intrinsically model dependent, as they rely on assumptions about the thermal history of the Universe, in particular the reheating temperature and the efficiency of thermal production mechanisms. Based on the ALP-photon coupling considerations in the previous section, the ALP–photon coupling in our scenarios is at least four orders of magnitude larger than the region where cosmological bounds become relevant. Consequently, the corresponding lifetime is extremely short on cosmological timescales, and the particle decays long before it could affect standard cosmological observables, rendering those constraints irrelevant for the parameter space we will explore.

Finally, turning our attention to collider searches, the most relevant  constraints are on the ALP–photon coupling and arise from channels with final states containing two or more photons~\cite{Brivio:2017ije,Bauer:2017ris,Bauer:2018uxu}. In particular, the dominant bounds are driven by LEP searches for associated ALP production of the type $e^+e^-\to\gamma a\to3\gamma$, by limits from exotic decays of on-shell $Z$ bosons, such as $Z\to\gamma a\to3\gamma$, and by LHC analyses targeting diphoton resonances or photon-fusion production of ALPs subsequently decaying into photon pairs. The present data constrain the region of the parameter space for ALP masses in $1\GeV\lesssim m_a\lesssim1\TeV$ and ALP-photon couplings that can be translated in our notation to be $f_a/c_{a\gamma\gamma}\gtrsim30\GeV$. This is actually in the same ballpark of values that limits the consistency of our description and therefore the existing bounds are automatically satisfied throughout the parameter space we are interested in.

Furthermore, when focussing on the future experimental facilities~\cite{Bauer:2018uxu}, the sensitivity to ALP-photon couplings increases through dedicated searches. Promising channels include associated production with photons, $Z$ or Higgs bosons at future $e^+e^-$ colliders such as FCC-ee, CEPC, ILC and CLIC, as well as diphoton resonance searches, photon-fusion production and exotic Higgs or $Z$ decays at the HL-LHC and possible future hadron colliders. Dedicated analyses of these signatures could therefore provide sensitivity to regions of our parameter space.

Moreover, additional constraints may arise from collider probes of ALP couplings to top quarks. Searches for heavy pseudoscalars decaying into $t\bar t$ pairs, as well as production in association with top quarks, provide bounds on the effective top coupling. For $m_a \gtrsim 400\GeV$, current analyses translate into limits of the order $|c_{tt}|/f_a \lesssim \cO(10^{-2}-10^{-3})\GeV^{-1}$, under specific assumptions on the ALP-gluon coupling and the ALP total decay rate~\cite{CMS:2019pzc,Esser:2023fdo,Anuar:2024qsz,Esser:2024pnc}. In the scenario considered here, the induced top coupling obtained after matching to the Yukawa basis satisfies $(\widehat{Y}'_u)^{33} \lesssim \cO(10^{-4})$, as discussed in App.~\ref{app:Running}, which lies well below the current experimental sensitivity. Consequently, collider observables involving top quarks do not impose relevant constraints on the parameter space explored in this work.\\

We will consider two different scenarios: i) an ALP that couples only to the first two generations and ii) an ALP that couples to the three families. Indeed, since the ALP--fermion couplings are proportional to the fermion masses, an ALP that has a tree-level (TL) coupling to $\tau$ leads to a noticeably different phenomenology compared to a $\tau$-phobic ALP. Besides this, we adopt specific benchmark flavour structures to analyse how the results depend on the underlying scenarios.

In what follows, in order to simplify the notation, we will redefine the coefficients $c_\ell^L$ and $c_e^R$ in Eq.~\eqref{eq:UVLagma} as $c^L$ and $c^R$ respectively. We recall that they account for the running RGE effects from the cut-off scale $\Lambda$ down to the ALP mass $m_a$.\footnote{The additional evolution down to the mass scale of the relevant charged lepton, scale at which the various observables are computed, is simply the purely SM RGE running and thus negligible.} We will better clarify the relation between these coefficients and the Lagrangian ones in the next section, after specifying the ALP-lepton coupling textures. 

\boldmath
\subsection{$e-\mu$ Framework}
\unboldmath

We first consider the simplified scenario of an ALP coupling only to the first two families, $e,\,\mu$. The leading-order (LO) contributions to the relevant observables, computed for an off-shell ALP in the limit $m_a\gg m_\mu$, are given by:
\begin{eqnarray}
\label{eq:ae}
    &&
    \begin{split}
    a_e^{\text{ALP}} \approx\dfrac{1}{8\pi^2}\dfrac{m_e m_\mu}{f_a^2}\Bigg\{&\dfrac{m_\mu^2}{m_a^2}
    \color{red}\Re[c_{\mu e}^L {c_{\mu e}^R}^*]\color{black}
    \left(-\dfrac{3}{2}+\log\left(\dfrac{m_a^2}{m_\mu^2}\right)\right)+\\
    &+\dfrac{\alpha_{\text{em}}}{\pi}\dfrac{m_e}{m_\mu}\color{blue}(c_{e e}^L-c_{e e}^R)\,\color{Green} c_{a\gamma\gamma}^{\text{eff}}\color{black} \left(-\dfrac{3}{2}+\log\left(\dfrac{\Lambda^2}{m_a^2}\right)\right)\Bigg\}\end{split}\\[2mm]
    &&
    \begin{split}
    \label{eq:amu}
    a_\mu^{\text{ALP}} \approx& 
    \dfrac{1}{8\pi^2}\dfrac{m_\mu^2}{f_a^2}\Bigg\{\dfrac{m_\mu^2}{m_a^2}
    \bigg[
    -\color{blue}\left(c_{\mu\mu}^L - c_{\mu\mu}^R\right)^2\color{black}\left( -\dfrac{11}{6} +  \log\!\left(\dfrac{m_a^2}{m_\mu^2} \right)\right)+\\
    &\hspace{2mm} +\dfrac{1}{6}\color{red}\big( |c_{\mu e}^L|^2 + |c_{\mu e}^R|^2 \big)\color{black}+\dfrac{m_e}{m_\mu}\color{red}\Re[c_{\mu e}^L {c_{\mu e}^R}^*]\color{black}
    \left(-\dfrac{13}{6}+\log\left(\dfrac{m_a^2}{m_e^2}\right)\right)
    \bigg]+\\
    &\hspace{2mm}+\dfrac{\alpha_{\text{em}}}{\pi}\color{blue}(c_{\mu\mu}^L-c_{\mu\mu}^R)\,\color{Green} c_{a\gamma\gamma}^{\text{eff}}\color{black} \left(-\dfrac{3}{2}+\log\left(\dfrac{\Lambda^2}{m_a^2}\right)\right)\Bigg\}\end{split}
    \end{eqnarray}
\begin{eqnarray}    &&
    \begin{split}
    \label{eq:br_mutoegamma}
    \mathcal{BR}(\mu\to e\gamma) \approx&
    \dfrac{\alpha_{\text{em}}}{1024\pi^4}\dfrac{m_\mu^5}{f_a^4 \Gamma_\mu}\color{red}\big(|c_{\mu e}^L|^2+|c_{\mu e}^R|^2\big)\color{black}\Bigg[
    \dfrac{m_\mu^2}{m_a^2}\,
    \color{blue}\big(c_{\mu\mu}^L - c_{\mu\mu}^R\big)\color{black}\,
    \left(-\dfrac{5}{3} +\log\!\left(\dfrac{m_a^2}{m_\mu^2} \right)\right)+\\
    &\hspace{4mm}-\dfrac{\alpha_{\text{em}}}{\pi}\color{Green} c_{a\gamma\gamma}^{\text{eff}}\color{black} \left(-\dfrac{3}{2}+\log\left(\dfrac{\Lambda^2}{m_a^2}\right)\right)\Bigg]^2
    \end{split}\\[2mm]
    &&
    \label{eq:br_mu3e}
    \mathcal{BR}(\mu\to 3e)\approx\dfrac{\alpha_{\text{em}}}{3\pi}\left(-\dfrac{11}{4}+\log\left(\dfrac{m_\mu^2}{m_e^2}\right)\right)\mathcal{B}\mathcal{R}(\mu\to e\gamma)\\[2mm]
    &&
    \label{eq:br_mue_conversion}
    \begin{split}
    \mathcal{BR}(\mu^- N \to e^- N) \approx&\dfrac{\xi}{512\pi^4}\dfrac{m_\mu^4}{f_a^4}\Bigg\{\dfrac{4}{3}\dfrac{m_\mu^4}{m_a^4}\color{blue}(c_{\mu\mu}^L-c_{\mu\mu}^R)^2\color{black}\left(-\dfrac{5}{3} +\log\left(\dfrac{m_a^2}{m_\mu^2} \right)\right)\times\\
    &\times\Bigg[\dfrac{5}{6}\color{red}\big(|c_{\mu e}^L|^2 + |c_{\mu e}^R|^2\big)\color{black}\left(-\dfrac{8}{5}+\log\left(\dfrac{m_a^2}{m_\mu^2}\right)\right)+ \\
    &\hspace{8mm}-\color{red}\Re\left[c_{\mu e}^L {c_{\mu e}^R}^*\right]\color{black}\left(-\dfrac{4}{3} +\log\left(\dfrac{m_a^2}{m_\mu^2}\right)\right) \Bigg]+\\
    &+\dfrac{\alpha_{\text{em}}^2}{\pi^2}\left(-\dfrac{3}{2}+\log\left(\dfrac{\Lambda^2}{m_a^2}\right)\right)^2\color{Green}\big(c_{a\gamma\gamma}^{\text{eff}}\big)^2\color{red}\big(|c_{\mu e}^L|^2 + |c_{\mu e}^R|^2\big)\color{black}\Bigg\}\,,\end{split}
\end{eqnarray}
where $\Gamma_\mu$ is the total muon decay rate and the prefactor $\xi$ is defined as
\be
    \xi=\dfrac{8\alpha_{\text{em}}^5 m_\mu Z_{\textit{eff}}^4 Z F_p^2}{\Gamma_{\textit{capt}}}\,,
\ee
being $\Gamma_{\textit{capt}}$ the total muon capture rate, $Z$ ($Z_{\textit{eff}}$) the (effective) atomic charge and $F_p$ the nuclear form factor. Explicit values for the Au and Al are \cite{Cornella:2019uxs}
\be
\xi_{\text{Au}} \approx 5.2\times 10^{12}\,,\qquad\qquad\xi_{\text{Al}}\approx 3.5\times 10^{12}\,.
\ee
Notice that in all these expressions, $\log$ stands for the natural logarithm.

The Wilson coefficients $c^L$ and $c^R$ are coloured in order to help guide the eye between the flavour conserving couplings in blue and the flavour violating ones in red. Although the effective coupling with photons $c^\text{eff}_{a\gamma\gamma}$ is not an independent coefficient, see Eq.~\eqref{ALPPhotonCouplingEFF}, a compact notation is used in these formulae and this coupling is shown in green to facilitate the identification of the photon penguin contributions.

A careful reader may notice that these expressions only contain ALP-couplings with the charged leptons and with photons, but not with a photon and a $Z$. The reason resides again in the fact that only the dominant contributions are shown in the expressions (and also in those for the three-family scenario reported below). In particular, when focussing on masses $m_a\gg m_\tau$, the form factors associated to the contributions involving $c_{a\gamma Z}^\text{eff}$ are extremely suppressed -- See the ancillary Mathematica file -- and therefore we have not shown them explicitly in the above formulae, but nevertheless we will consider them in our numerical analysis.

\subsection{Three-Family Framework}

In the case of an ALP coupling to three generations, the ALP-couplings to $\tau$ play a dominant role. Not only should the $\tau$ phenomenology be considered, but also the expressions for the $\mu$ and electron processes introduced above become substantially modified. All in all, ignoring the specific hierarchies among the ALP-lepton couplings, the expressions of the relevant observables simplify as follows: for an off-shell ALP in the limit $m_a\gg m_\tau$,
\begin{eqnarray}
    &&
    \begin{split}
    \label{eq:g2electrontau}
    a_e^{\text{ALP}} \approx&
    \dfrac{1}{8\pi^2} \dfrac{m_e m_\tau}{f_a^2}\Bigg\{\dfrac{m_{\tau}^2}{m_a^2}\color{red}\Re[c_{\tau e}^L {c_{\tau e}^R}^*]\color{black}\left(-\dfrac{3}{2}+\log\left(\dfrac{m_a^2}{m_\tau^2} \right)\right)+\\
    &\hspace{22mm}+\dfrac{m_e}{m_\tau}\Bigg[\dfrac{\alpha_{\text{em}}}{\pi}\color{blue}(c_{e e}^L-c_{e e}^R)\color{Green} c_{a\gamma\gamma}^{\text{eff}}\color{black}\left(-\dfrac{3}{2}+\log\left(\dfrac{\Lambda^2}{m_a^2}\right)\right)\Bigg]\Bigg\}\end{split}\\[2mm]
    &&
    \begin{split}
    \label{eq:g2muontau}
    a_\mu^{\text{ALP}} \approx&
    \dfrac{1}{8\pi^2} \dfrac{m_\mu m_\tau}{f_a^2}\Bigg\{\dfrac{m_{\tau}^2}{m_a^2}\color{red}\Re[c_{\tau \mu}^L {c_{\tau \mu}^R}^*]\color{black}\left(-\dfrac{3}{2}+\log\left(\dfrac{m_a^2}{m_\tau^2} \right)\right)+\\
    &\hspace{22mm}+\dfrac{m_\mu}{m_\tau}\Bigg[\dfrac{\alpha_{\text{em}}}{\pi}\color{blue}(c_{\mu\mu}^L-c_{\mu\mu}^R)\color{Green} c_{a\gamma\gamma}^{\text{eff}}\color{black}\left(-\dfrac{3}{2}+\log\left(\dfrac{\Lambda^2}{m_a^2}\right)\right)\Bigg]\Bigg\}\end{split}\\[2mm]
    &&
    \begin{split}
    \label{eq:muegammatau}
    \mathcal{B}\mathcal{R}(\mu\to e\gamma)\approx&\dfrac{\alpha_{\text{em}}}{1024\pi^4}\dfrac{m_{\mu}^3 m_{\tau}^2}{f_a^4\Gamma_{\mu}}\Bigg\{\dfrac{m_{\tau}^4}{m_a^4}\color{red}\big(|c_{\tau e}^L|^2 |c_{\tau\mu}^R|^2+|c_{\tau e}^R|^2|c_{\tau\mu}^L|^2\big)\color{black}\left(-\dfrac{3}{2}+\log\left(\dfrac{m_a^2}{m_{\tau}^2}\right)\right)^2+\\
    &+\dfrac{\alpha_{\text{em}}^2}{\pi^2}\dfrac{m_\mu^2}{m_\tau^2}\color{Green} (c_{a\gamma\gamma}^{\text{eff}})^2\color{black}\color{red}\big(|c_{\mu e}^L|^2+|c_{\mu e}^R|^2\big)\color{black}\left(-\dfrac{3}{2}+\log\left(\dfrac{\Lambda^2}{m_a^2}\right)\right)^2\Bigg\}\end{split}\\[2mm]
    &&
    \begin{split}
    \label{eq:tauegamma}
    \mathcal{BR}(\tau\to e\gamma) \approx&
    \dfrac{\alpha_{\text{em}}}{1024\pi^4}\dfrac{m_\tau^5}{f_a^4 \Gamma_\tau}\color{red}\big(|c_{\tau e}^L|^2+|c_{\tau e}^R|^2\big)\color{black}\Bigg[
    \dfrac{m_\tau^2}{m_a^2}\,
    \color{blue}\big(c_{\tau\tau}^L - c_{\tau\tau}^R\big)\color{black}\,
    \left(-\dfrac{5}{3} +\log\!\left(\dfrac{m_a^2}{m_\tau^2} \right)\right)+\\
    &\hspace{4mm}-\dfrac{\alpha_{\text{em}}}{\pi}\color{Green} c_{a\gamma\gamma}^{\text{eff}}\color{black} \left(-\dfrac{3}{2}+\log\left(\dfrac{\Lambda^2}{m_a^2}\right)\right)\Bigg]^2\end{split}\\[2mm]
    &&
    \label{eq:taumugamma}
    \mathcal{BR}(\tau\to \mu\gamma) \approx\color{red}\dfrac{\big(|c_{\tau \mu}^L|^2+|c_{\tau \mu}^R|^2\big)}{\big(|c_{\tau e}^L|^2+|c_{\tau e}^R|^2\big)}\color{black}\mathcal{BR}(\tau\to e\gamma)\\[2mm]
    &&
    \label{eq:mueconversiontau}
    \mathcal{BR}(\mu^- N \to e^- N)\approx \dfrac{2\,\xi}{\alpha_{\text{em}}}\dfrac{\Gamma_\mu}{m_\mu}\mathcal{B}\mathcal{R}(\mu\to e\gamma)\\[2mm]
    &&
    \label{eq:mu3etau}
    \mathcal{BR}(\mu\to3e)\approx\dfrac{\alpha_{\text{em}}}{3\pi}\left(-\dfrac{11}{4}+\log\left(\dfrac{m_\mu^2}{m_e^2}\right)\right)\mathcal{B}\mathcal{R}(\mu\to e\gamma)\\[2mm]
    &&
    \label{eq:tau3e}
    \mathcal{BR}(\tau\to 3e) \approx\dfrac{\alpha_{\text{em}}}{3\pi}\left(-\dfrac{11}{4}+\log\left(\dfrac{m_\tau^2}{m_e^2}\right)\right)\mathcal{B}\mathcal{R}(\tau\to e\gamma)\\[2mm]
    &&
    \label{eq:tau3mu}
    \mathcal{BR}(\tau\to 3\mu)\approx\dfrac{\alpha_{\text{em}}}{3\pi}\left(-\dfrac{11}{4}+\log\left(\dfrac{m_\tau^2}{m_\mu^2}\right)\right)\mathcal{B}\mathcal{R}(\tau\to \mu\gamma)
\end{eqnarray}
\begin{eqnarray}
    &&
    \begin{split}
    \label{eq:tautomuminee}
    \mathcal{BR}(\tau^- \to \mu^- &e^+ e^-) \approx\dfrac{\alpha_{\text{em}}^2}{3072\pi^5}\dfrac{m_\tau^5}{f_a^4 \Gamma_\tau}\color{red}\left(|c_{\tau\mu}^L|^2+|c_{\tau\mu}^R|^2\right)\color{black}\times\\
    &\times\Bigg\{\dfrac{m_{\tau}^4}{m_a^4}\color{blue}\left(c_{\tau \tau}^L-c_{\tau\tau}^R\right)^2\color{black}\Bigg[\dfrac{13}{36}\left(-\dfrac{64}{39}+\log\left(\dfrac{m_a^2}{m_\tau^2}\right)\right)\left(-\dfrac{4}{3}+\log\left(\dfrac{m_a^2}{m_\tau^2}\right)\right)+\\
    &\hspace{23mm}+\left(-\dfrac{5}{3}+\log\left(\dfrac{m_a^2}{m_\tau^2}\right)\right)^2\left(-3+\log\left(\dfrac{m_\tau^2}{m_e^2}\right)\right)\Bigg]+\\
    &\hspace{7mm}+\dfrac{\alpha_{\text{em}}^2}{\pi^2}\color{Green}\big(c_{a\gamma\gamma}^{\text{eff}}\big)^2\color{black}\left(-\dfrac{3}{2}+\log\left(\dfrac{m_a^2}{m_\tau^2}\right)\right)^2\left(-3+\log\left(\dfrac{m_\tau^2}{m_e^2}\right)\right)\Bigg\}
    \end{split}\\[2mm]
    &&
    \begin{split}
    \label{eq:tautoeminmumu}
    \mathcal{BR}(\tau^- \to e^- &\mu^+ \mu^-) \approx\dfrac{\alpha_{\text{em}}^2}{3072\pi^5}\dfrac{m_\tau^5}{f_a^4 \Gamma_\tau}\color{red}\left(|c_{\tau e}^L|^2+|c_{\tau e}^R|^2\right)\color{black}\times\\
    &\times\Bigg\{\dfrac{m_{\tau}^4}{m_a^4}\color{blue}\left(c_{\tau \tau}^L-c_{\tau\tau}^R\right)^2\color{black}\Bigg[\dfrac{13}{36}\left(-\dfrac{64}{39}+\log\left(\dfrac{m_a^2}{m_\tau^2}\right)\right)\left(-\dfrac{4}{3}+\log\left(\dfrac{m_a^2}{m_\tau^2}\right)\right)+\\
    &\hspace{23mm}+\left(-\dfrac{5}{3}+\log\left(\dfrac{m_a^2}{m_\tau^2}\right)\right)^2\left(-3+\log\left(\dfrac{m_\tau^2}{m_\mu^2}\right)\right)\Bigg]+\\
    &\hspace{7mm}+\dfrac{\alpha_{\text{em}}^2}{\pi^2}\color{Green}\big(c_{a\gamma\gamma}^{\text{eff}}\big)^2\color{black}\left(-\dfrac{3}{2}+\log\left(\dfrac{m_a^2}{m_\tau^2}\right)\right)^2\left(-3+\log\left(\dfrac{m_\tau^2}{m_\mu^2}\right)\right)\Bigg\}
    \end{split}
\end{eqnarray}
In the above expressions, $\Gamma_\tau$ stands for the total $\tau$ decay rate. The attributes of the ALP couplings mentioned in the two-family case also apply here. 

A few more comments are in order. The expressions in Eqs.~\eqref{eq:tautomuminee} and \eqref{eq:tautoeminmumu} only show the one-loop contributions proportional to $\alpha^2_\text{em}$. These are indeed the dominant contributions especially for large $m_a$ values. Nevertheless, the tree level contributions and the interference ones are in general not negligible and are taken into consideration in our numerical analysis. On the other hand, our analysis does not include the processes $\tau^-\to \mu^+ e^- e^-$ and $\tau^-\to e^+ \mu^- \mu^-$ since a dedicated analytical study beyond the present literature would be necessary, a task which we leave for a separate work. Nevertheless, their impact in our numerical results is expected to be negligible given the present and future experimental sensitivities.

Last but not least, we point out that for certain flavour textures for which the ALP-$\tau$ couplings are not negligible yet still small relative to other ALP-lepton couplings, the complete expressions for the above observables should be used. Indeed, due to the enhancement of the $\tau$-mass it may occur that contributions proportional to ``small'' ALP-$\tau$ couplings are as important as those involving ``large'' ALP-$\mu$ couplings. These expressions can be found in the ancillary Mathematica file.

\subsection{Benchmark Flavour Textures}
\label{sec:flavour_textures}

In order to explore the richness of the parameter space, we focus on a set of flavour textures for the ALP--lepton couplings that allow us to extract relevant information for each case. 
We will first focus on the two-family case and then move to the full three-family one. There are two competing ingredients that determine the expected results in each case: the dominant couplings present in the different observables and the present (and/or future) experimental sensitivities of these observables. The discussion that follows will be driven by both in order to identify flavour structures that lead to different phenomenological regions in the parameter space.

Before going into details of the different textures, it is necessary to specify the scale at which they are defined and the connection with the Lagrangian coefficients $c_\psi$. The working hypothesis is that the high-energy Lagrangian in Eq.~\eqref{eq:UVLag} at the scale $\Lambda$ is defined in the Minimal Lepton Flavour Violation (MLFV)~\cite{DAmbrosio:2002vsn,Cirigliano:2005ck,Davidson:2006bd,Gavela:2009cd,Alonso:2011jd,Dinh:2017smk,Alonso:2012fy,Alonso:2013nca,Alonso:2013mca} flavour basis with the charged lepton Yukawa diagonal~\footnote{Without attempting to identify and construct an explicit UV completion that yields this condition, one may envisage a scenario consistent with the MLFV framework in which the flavour symmetry is broken at a scale $\Lambda'>\Lambda$. In such a setup, a non-trivial ALP–lepton flavour structure can emerge from the action of heavy mediators that remain dynamical down to the scale $\Lambda$, where the PQ symmetry is subsequently broken.} or any other flavour model where the three lepton generations are ruled by a single symmetry group (e.g.~\cite{Blankenburg:2012nx,Feldmann:2016hvo,Alonso:2016onw,Arias-Aragon:2020bzy}). The neutrino sector discussion depends on the specific mechanism assumed to generate masses for the active neutrinos. For example, within the minimal extension of the SM involving the Weinberg operator~\cite{Weinberg:1979sa}, a consistent MLFV treatment would be to consider the PMNS matrix fully contained in the Wilson coefficient of the Weinberg operator. It follows that the relation between the coefficients $c_\psi$ and $c'_\psi$ is instead highly simplified and  $c_\ell=c'_\ell$ and $c_e=c'_e$.

The RGE running introduces corrections to these couplings moving from $\Lambda$ to the scale of the ALP mass $m_a$. These effects have been studied at length in the literature~\cite{Chala:2020wvs,Bauer:2020jbp,Bonilla:2021ufe} and we avoid reporting here the explicit expressions of the relevant $\beta$-functions. Instead, we use the \alpaca\ tool~\cite{Alda:2025nsz} in order to correctly take into account the RGE effects and conclude that the following textures are stable under the running, i.e.\ the quantum corrections turn out smaller or at most of the same order of the present entries. This is due to a combination of two main factors: i) tree-level couplings with gluons or quarks are vanishing by construction at the scale $\Lambda$; ii) the suppression of the electron and muon masses with respect to the tau mass. It follows that the tau-sector receives the largest corrections, but they have no impact in our analysis. We will come back to this aspect with a more numerical estimate at the end of each subsection. 

\boldmath
\subsection*{$e-\mu$ Framework}
\unboldmath

When the ALP only couples to the first two generations, the formulae in Eqs.~\eqref{eq:ae}--\eqref{eq:br_mue_conversion} may provide some insights into simplifying the multi-dimensional parameter space. A warning message is in order: these expressions only show the leading contributions, while subleading ones may contain other dependencies on the ALP--lepton couplings. The discussion that follows mainly refers to the dominant contributions and we will comment on the subleading ones when necessary. 

First of all, the RH and LH couplings, $c^{R}$ and $c^{L}$, typically enter expressions for the observables through combinations that are (anti-)symmetric under the exchange of $c^{R}$ and $c^{L}$. Consequently, switching off one of the two chiralities or introducing a strong hierarchy between them leads to the same qualitative behaviour as assuming identical flavour structures for both. For this reason, we restrict our analysis to the latter scenario.

There are, however, notable exceptions. The magnetic dipole moments of the electron and the muon, as well as the $\mNeN$ decay, contain terms proportional to the product $c^{R} c^{L}$, and therefore are sensitive to the relative size of the two chiralities. Nevertheless, assuming a similar flavour structure for $c^{R}$ and $c^{L}$ does not limit the generality of our conclusions: a large hierarchy between the two can in general be reabsorbed into a redefinition of the scale $f_{a}$, yielding a phenomenologically equivalent setup to the case with $c^{R}$ and $c^{L}$ of comparable magnitude.

Focussing now on the internal flavour structure of the couplings, we can discuss the impact of the diagonal and off-diagonal entries. Most expressions present the product of the two kinds of entries in some combination, that is $c^{L/R}_{ii}\times c^{L/R}_{ij}$. The magnetic dipole moments of both electron and muon show other type of dependencies: they depend on either a product of only off-diagonal entries or alternatively a combination of only diagonal entries, explicitly or through the effective ALP-coupling with photons that is tantamount to saying the difference between the flavour conserving chiral couplings. The simultaneous presence of different combinations may translate into a more difficult understanding of the numerical results with respect to the analytical dependence of the observables from the ALP couplings.

All in all, we can identify three flavour structures that present different phenomenologies. The first is a fully diagonal structure, which would imply that any flavour changing process is extremely suppressed, as indeed the expressions in Eqs.~\eqref{eq:br_mutoegamma}--\eqref{eq:br_mue_conversion} identically vanish. The only relevant observables are the magnetic dipole moments of electron and muon that receive contributions from the photon penguin diagram.

If, instead, we consider a texture with only off-diagonal entries, or with off-diagonal entries very much dominant with respect to the diagonal ones, one may expect a completely different phenomenological scenario. However, this is not the case as the leading contributions of the flavour changing processes are suppressed with respect to those of the magnetic dipole moments. All in all, these two extreme scenarios give rise to very similar constraints on the parameter space. We thus consider the following texture as a benchmark case:
\begin{equation}
\label{eq:texture_2a}
\text{Texture 2a):}\qquad    
c^{L/R}\sim \left( \begin{array}{cc}
        0 &\mathcal{O}(1)\\
        \mathcal{O}(1) &  0\\
    \end{array}\right)\,.
\end{equation}

A more democratic texture with all the entries of the same order of magnitude leads, instead, to a much richer phenomenology, where the flavour changing processes play a relevant role considering the present constraints and the future expected sensitivities: 
\begin{equation}
\label{eq:texture_2b}
\text{Texture 2b):}\qquad 
c^{L/R}\sim \left( \begin{array}{cc}
        \mathcal{O}(1) &\mathcal{O}(1)\\
        \mathcal{O}(1) & \mathcal{O}(1)\\
    \end{array}\right)\,.
 \end{equation}

These two textures are stable under the running between the scales we consider in our analysis. To illustrate this point, we show in App.~\ref{app:Running} the explicit result of assuming these textures at the high-scale $\Lambda=10\TeV$ and applying the RGE running down to any scale of the ALP mass down to $m_a=50\GeV$, using \alpaca. In the $e-\mu$ sector, there is no relevant correction that modifies these initial structures: the running contributes to the flavour conserving $\tau$-entry with a value of $\sim10^{-6}$ in the best scenario. The latter may represent a dangerous outcome as the $\tau$-contributions are generically enhanced by $m_\tau/m_\mu$ or $m_\tau/m_e$ to some powers, but a closer look to the expressions in Eqs.~\eqref{eq:g2electrontau}--\eqref{eq:tautoeminmumu} for the three-family framework reveals that the $c^{L,R}_{\tau\tau}$ couplings always appear proportional to the flavour violating $\tau$-entries, $c^{L,R}_{e\tau}$ or $c^{L,R}_{\mu\tau}$, which receive contributions from the running that are even more suppressed.~\footnote{The one-loop contributions to these flavour violating entries are suppressed by the neutrino masses and are thus negligible. The subsequent contributions appear at two-loop level and thus also very much suppressed.} As a consequence, these contributions are safely subleading with respect to those in Eqs.~\eqref{eq:ae}--\eqref{eq:br_mue_conversion} involving exclusively the ALP-lepton couplings with the first two generations. 

In App.~\ref{app:Running}, we also report the dominant contributions to the low-energy hadronic couplings. These correspond to the flavour-conserving interactions of the ALP with the top and bottom quarks and are sufficiently suppressed to produce no appreciable impact on hadronic observables. This can be verified by combining the effective analysis of Eq.~\eqref{EffectiveLag4FermionOperators} with the numerical results presented in Tab.~\ref{tab:Running2Families} and implementing them in \texttt{flavio}~\cite{Straub:2018kue}. While this discussion does not provide an exhaustive treatment of hadronic phenomenology, as for instance it does not include observables induced at one loop, it nonetheless offers a sound justification for neglecting hadronic constraints when considering these textures.

\subsection*{Three-Family Framework}

Once the ALP couples to the three lepton generations, more observables can be considered and the relevant expressions are Eqs.~\eqref{eq:g2electrontau}--\eqref{eq:tautoeminmumu}. As for the previous case, the couplings $c^R$ and $c^L$ are both present, and the case of hierarchies among them provides the same qualitative phenomenology as the case with similar magnitudes for many observables. Indeed, although considering the extreme case where either $c^R=0$ or $c^L=0$, leading to several contributions to vanish, the presence of the photon penguin terms implies just a $\sim 10^{-2}$ suppression with respect to the case with both sizeable couplings. This suppression is generically easily reabsorbed into a redefinition of the scale $f_a$.

If we now focus on the flavour structure of these textures, the flavour diagonal couplings enter either explicitly or through the effective ALP-photon coupling. Although the latter are suppressed by powers of $\alpha_\text{em}$, they turn out to be dominant for large $m_a$ values. The reason can be easily appreciated for example looking closer to Eq.~\eqref{eq:g2electrontau}. The purely axion-mediated contribution in the first line is proportional to $m_\tau^2/m_a^2$, while the photon penguin in the second line is proportional to  $\alpha_\text{em} m_e/(\pi m_\tau)$ with the $m_a$-dependence entering only in the logarithm: the latter contribution mildly varies with $m_a$, while the first scales quadratically with $m_a$.

It is then interesting to analyse to which extent the off-diagonal entries provide specific phenomenological signatures. One benchmark texture that we will consider is 
\begin{equation}
\label{eq:texture_3a}
\text{Texture 3a):}\qquad\qquad  
    c^{L/R}\sim \left( \begin{array}{ccc}
        \mathcal{O}(1) &\mathcal{O}(1) &\mathcal{O}(1) \\
        \mathcal{O}(1) & \mathcal{O}(1)&\mathcal{O}(1)\\
        \mathcal{O}(1) & \mathcal{O}(1)&\mathcal{O}(1)
    \end{array}\right)
\end{equation}
that represents a completely democratic scenario. The elements in the diagonal are taken $\mathcal{O}(1)$, but smaller values would lead to the same phenomenological results, according to the previous discussion.

Modifying the previous texture to suppress the couplings with the $\tau$ would converge towards texture 2b), as expected. Textures of the type 
\begin{equation}
\label{eq:texture_3b}
\text{Texture 3b):}\qquad\qquad 
    c^{L/R}\sim \left( \begin{array}{ccc}
        \mathcal{O}(1) &
        \mathcal{O}(1)& 
        \lesssim\mathcal{O}(10^{-3}) \\
        \mathcal{O}(1) & 
        \mathcal{O}(1)&
        \lesssim\mathcal{O}(10^{-3})\\
        \lesssim\mathcal{O}(10^{-3})&
        \lesssim\mathcal{O}(10^{-3})&
        \lesssim\mathcal{O}(10^{-3})
    \end{array}\right)
\end{equation}
will then be discussed together with Texture 2b) and referred to as $\tau$-phobic cases. 

A third type of flavour benchmark texture that may be considered is the $e$-phobic case, where the $\mu-\tau$ sector has the largest couplings with the ALP:
\begin{equation}
\label{eq:texture_3c}
\text{Texture 3c):}\qquad\qquad  
     c^{L/R}\sim \left( \begin{array}{ccc}
        \lesssim\mathcal{O}(10^{-4}) &
        \lesssim\mathcal{O}(10^{-4})&
        \lesssim\mathcal{O}(10^{-4}) \\
        \lesssim\mathcal{O}(10^{-4}) & 
        \mathcal{O}(1)&
        \mathcal{O}(1)\\
        \lesssim\mathcal{O}(10^{-4})&
        \mathcal{O}(1)&
        \mathcal{O}(1)
    \end{array}\right)\,.
\end{equation}
In this case all contributions to the observables that involve an electron are then extremely suppressed, and the only relevant observables to the analysis are expected to be the $a_\mu^\text{ALP}$, the $\tmg$, $\ttm$ and $\tau^-\to \mu^- e^+ e^-$ decays. These observables have however lower sensitivities and therefore the analysis is less constraining.

The last trivial modification of texture 3a) is when the muon sector is suppressed, that is a $\mu$-phobic scenario, corresponding to
\begin{equation}
\label{eq:texture_3d}
\text{Texture 3d):}\qquad\qquad 
     c^{L/R}\sim \left( \begin{array}{ccc}
        \mathcal{O}(1) &
        \lesssim\mathcal{O}(10^{-4})&
        \mathcal{O}(1) \\
        \lesssim\mathcal{O}(10^{-4}) & 
        \lesssim\mathcal{O}(10^{-4})&
        \lesssim\mathcal{O}(10^{-4})\\
        \mathcal{O}(1)&
        \lesssim\mathcal{O}(10^{-4})&
        \mathcal{O}(1)
    \end{array}\right)\,.
\end{equation}
This is complementary to the previous case, where contributions to observables involving a muon are extremely suppressed, and the relevant processes are expected to be $a_e^\text{ALP}$, the $\teg$, $\tte$ and $\tau^-\to e^- \mu^+ \mu^-$ decays. Besides the muon magnetic dipole moment, the other observables are not well known and their impact on the numerical analysis should be weak. 

Anticipating the results from the numerical analysis for textures 3c) and 3d), the expected subleading contributions to $\meg$ and $\mNeN$ are not so negligible and have an impact in the plots, although in many cases not so evident.

As for the 2-family case, these coupling matrices are assumed at the high-scale $\Lambda$ and one should carefully check their stability under the RGE running from that scale down to the ALP mass. By adopting the \alpaca\ code, we have done this exercise and found out that the largest contributions are smaller or at most equal to the present entries. Moreover, once again, we see that the induced ALP-couplings to third-generation quarks are tiny, thus leading to a negligible hadron phenomenology. App.~\ref{app:Running} contains the corresponding discussion and shows the explicit results with $\Lambda=10\TeV$ and $m_a\in \left[50,400\right]\GeV$.

\section{Phenomenological Analysis}
\label{sec:results}

We have performed a $\chi^2$ analysis of all relevant observables at the $3\sigma$ level for the different coupling textures, taking into account the observables listed in the previous section. Given that the two determinations of the $(g-2)_e$ are not mutually consistent, we begin with a conservative approach based on the Rubidium determination, and subsequently address the anomaly implied by the Caesium result.

Considering the number of parameters involved, $m_a$, $\Lambda$, $c^{L,R}_{ij}$, we assume real Wilson coefficients to simplify the analysis and moreover we do not present our results in terms of a global fit. Instead, we marginalise with respect to two parameters at a time, fixing the others to specific values. The aim is to illustrate specific features of the textures with two-dimensional plots where the allowed parameter space accounts for the present experimental constraints and the future prospects.

In doing that, we avoid {\it ad hoc} cancellations among different contributions, by assigning specific values and signs to the fixed parameters in each plot. For example, the flavour conserving couplings enter most of the time as a difference between the LH and RH ones, $(c^L_{ii}-c^R_{ii})$: to avoid these terms to vanish, we generically assume these coefficients to be equal in modulus but with opposite sign, $c^L_{ii}=-c^R_{ii}$.

Moreover, as is common in an EFT analysis, bounds do not apply individually on the Lagrangian coefficients or on the scale, that is $c^{L,R}$ and $f_a$ or $m_a$ in our case, but on their ratios, namely $c^{L,R}/f_a$ or $c^{L,R}/m_a$. This is why most analyses are done by fixing the value of the scale and investigating the coefficient parameter space, or, on the contrary, by fixing the coefficients to specific values and studying the bounds on the scale. We will follow both approaches.

\begin{figure}[h!]
\centering
\includegraphics[width=0.455\textwidth]{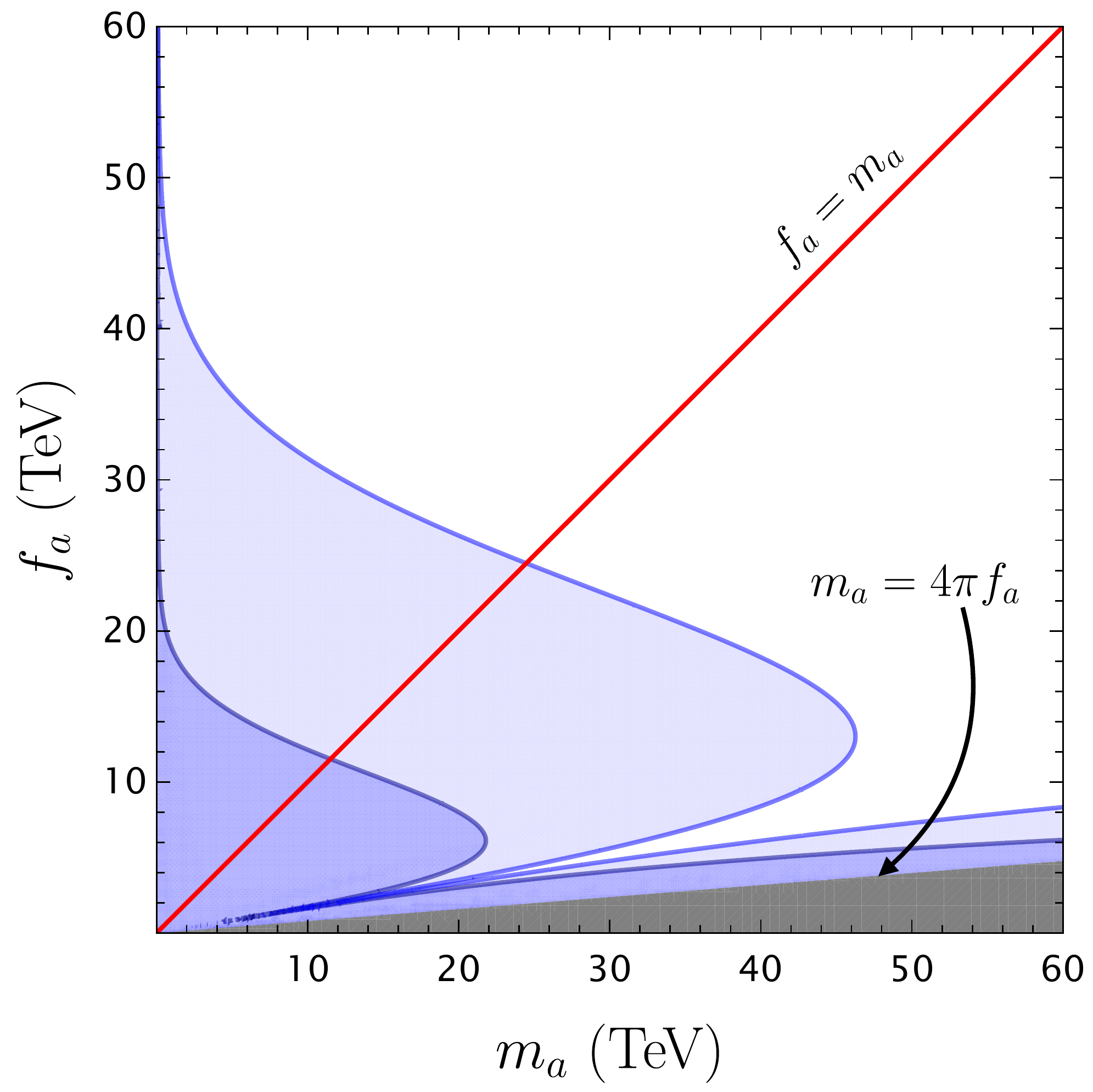}
\hfill
\includegraphics[width=0.45\textwidth]{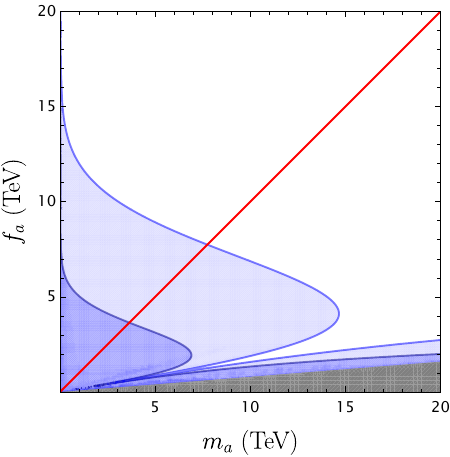}
\caption{\it $m_a\times f_a$ parameter space for the texture 3a), having fixed all the Lagrangian coefficients to $c^L=-c^R=1$ (left) and to $c^L_{ii}=-c^R_{ii}=1$ and $c^L_{ij}=-c^R_{ij}=0.1$ (right). The darker (lighter) blue region denotes the parameter space excluded by present experimental constraints (future prospects), at $3\sigma$. The red line represents the condition $m_a=f_a$. The gray region is excluded by the theoretical consistency condition $m_a<4\pi f_a$.}
 \label{Labelmafa12fp}
\end{figure}

Before proceeding with the discussion of the textures, we present a study of the $m_a\times f_a$ parameter space, fixing all the coefficients as discussed above. The result can be seen in Fig.~\ref{Labelmafa12fp}. The plot on the left is obtained in the specific case of the texture 3a), by fixing the coefficients as $c^L=-c^R=1$ both for the flavour conserving and violating entries. The darker blue region is excluded at $3\sigma$ considering the present experimental bounds, while the lighter one also includes future prospects. The grey region instead is excluded due to a theoretical consistency condition: $m_a<4\pi f_a$, assuming in this case the cut-off to be $\Lambda=4\pi f_a$, which would reflect a strongly interacting theory. The red line refers to the condition $m_a=f_a$ and  only serves to divide the two regions of the parameter space, where $m_a$ is larger or smaller than $f_a$. 

The processes that dominate the fit are $\mu \to e \gamma$ for the present available data, and $\mu N \to e N$ for the future prospects. 
As expected from the previous expressions of these observables, we can appreciate the effect of two competing tendencies. On the one hand, the contributions to the photon penguin are dominant over almost the entire parameter space except for the spike, where they (partially) vanish. Their dependence on the ALP mass appears exclusively in the logarithmic terms and is suppressed by $\left(\alpha_\text{em}/f_a^2\right)^n$. On the other hand, the direct charged lepton contributions, which are proportional to $1/(m_a f_a)^{2n}$, are relevant only in the spike region and for low values of $m_a$. All in all, the unconstrained parameter space is very large and {\it there is no experimental preference for the allowed white regions with $m_a<f_a$ or with $m_a>f_a$.}  The latter is only further constrained by the theoretical consistency limit that starts to dominate for $m_a\gtrsim 200\TeV$.
The viability of this region of the parameter space is typically overlooked by the prejudice of assuming an ALP with the same hierarchical relation between its mass and the scale $f_a$ as for the traditional QCD axion, i.e.\ $m_a\ll f_a$.

The plot on the right closely resembles the one on the left, the only difference being the choice of flavour violating couplings, which are now fixed to $c^L_{ij} = -c^R_{ij} = 0.1$. Although this setup no longer corresponds to the texture 3a), it is nevertheless useful to illustrate how the exclusion region depends on the overall magnitude of the flavour violating entries. Since the dominant observables scale as $|c^L_{ij}|^2 |c^R_{ij}|^2$, lowering $c^{L,R}_{ij}$ by one order of magnitude suppresses the exclusion bounds by a factor of 3. More importantly, this modification does not alter the shape of the excluded region; it simply produces a uniform rescaling in the $m_a \times f_a$ plane. However, for textures with even smaller off-diagonal entries, the relative relevance among the observables may completely change. Indeed, observables with a stronger dependence on the flavour conserving charged lepton couplings may provide the strongest constraints. For example, certain tau-lepton decay modes may also become relevant through their dependence on $c^{L,R}_{\tau\tau}$, although their constraints remain systematically weaker than those from $\mu \to e \gamma$, which continues to provide the dominant bound. At the same time, such a strong suppression would push the exclusion limit of both $f_a$ and $m_a$ to very low scales, entering in the region where our description breaks down and thus no conclusion can be deduced. 

All in all, in view of the above considerations, we refrain from performing a systematic scan over the coupling values. Instead, we concentrate on textures with genuinely distinct internal structures that cannot be mapped onto each other through a mere rescaling of their entries.

\begin{figure}[h!]
\centering
\includegraphics[width=0.47\textwidth]{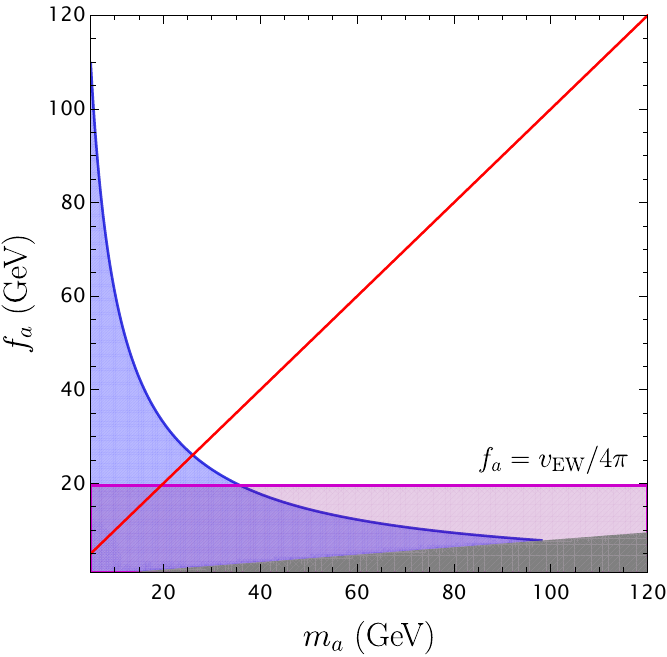}
\hfill
\includegraphics[width=0.457\textwidth]{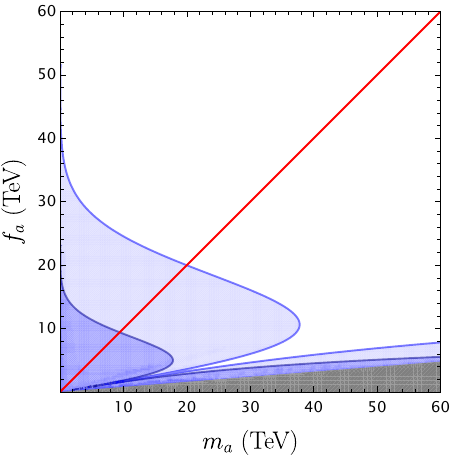}
\caption{\it $m_a\times f_a$ parameter space for the textures 2a) on the left and 2b) on the right, having fixed all the Lagrangian coefficients to $c^L=-c^R=1$. The darker (lighter) blue region denotes the parameter space excluded by present experimental constraints (future prospects), at $3\sigma$. The red line represents the condition $m_a=f_a$. The gray and purple regions are excluded by the theoretical consistency conditions $m_a<4\pi f_a$ and $f_a< v_{\text{EW}}/4\pi$, respectively.}
\label{Labelmafa12fpTexture2}
\end{figure}

Fig.~\ref{Labelmafa12fpTexture2} illustrates the same parameter space as in Fig.~\ref{Labelmafa12fp}, but for Textures 2a) (left) and 2b) (right). First of all, comparing the four plots, we can notice that the experimental bounds are much weaker in the case of Texture 2a), presenting even a different profile shape. The difference in scales arises from the contributions of $c_{a\gamma\gamma}^{\text{eff}}$, which are absent in Texture 2a). Texture 2b) and Texture 3a) present a similar scale, being the relevant difference only the contribution from $c_{\tau\tau}$ inside $c_{a\gamma\gamma}^{\text{eff}}$. In the regime of the spike where the explicit charged lepton contribution is relevant, we can appreciate the effect of the $\tau$ mass enhancement: when $\tau$ couplings are active, the corresponding contributions are enhanced by $m_\tau/m_\mu$ terms and, correspondingly, the values of $m_a$ and $f_a$ have to be larger in order to saturate the $\mu\to e\gamma$ and $\mNeN$ constraints that dominates the experimental bounds.

More into details, in the plot on the left for Texture 2a), the present bounds are mainly given by the anomalous $g-2$ of the muon and the future prospects from charged lepton decays are not visible as they are weaker. For this texture, the induced ALP-photon coupling is extremely suppressed and thus the photon penguin does not provide interesting contributions. We remain with the pure charged lepton contributions and their $1/(m_a f_a)^4$ dependence. Instead, in the plot on the right for Texture 2b), the present constraints arise from $\meg$ decay and the future prospects from $\mNeN$, and in particular the effective photon coupling is sizeable and consequently the photon penguin contribution is dominant. 

For Texture 2a) there are values of the ALP mass for which the values of $f_a$ already fall below the scale of validity of our description, $f_a\sim30\GeV$. Although this plot strictly holds for $c^L_{\mu e}=1=-c^R_{\mu e}$, slight changes to these values do not lead to different results, and therefore we expect it to be unbounded from the lepton observables considered. Texture 2b), instead, is perfectly consistent with our EFT treatment. As for the previously discussed Texture 3a), there is no experimental preference for any of the two regions separated by the red line. \\

In what follows, for each texture, we will include plots that illustrate the possible existing correlations between different ALP-lepton couplings: in most of the cases, the behaviour that we can naively infer from the formulae associated to the most relevant observables can be easily identified in the plots. However, in some cases the expected leading contributions are not so dominant, and/or observables with weak sensitivities also play a relevant role in the $\chi^2$ analysis and consequently the understanding of the parameter space is more involved. Moreover, we will always focus on comparing $m_a<f_a$ and $m_a>f_a$ to underline that both regions deserve to be equally studied.

\subsection{Analysis of the Textures}

\subsubsection*{Texture 2a)}

Texture 2a) only presents flavour violating entries and the relevant observables are the $g-2$ of the electron and muon. Indeed, the muon decays and the $\mNeN$ conversion are suppressed, as they depend on the product of flavour conserving and violating couplings. As the plot on the left of Fig.~\ref{Labelmafa12fpTexture2} shows, the scales of $m_a$ and $f_a$ corresponding to the present experimental bounds are extremely low and in part violate the consistency limit of our description,  $f_a>v_\text{EW}/4\pi$. 

\begin{figure}[h!]
\centering
\includegraphics[width=1.05\textwidth]{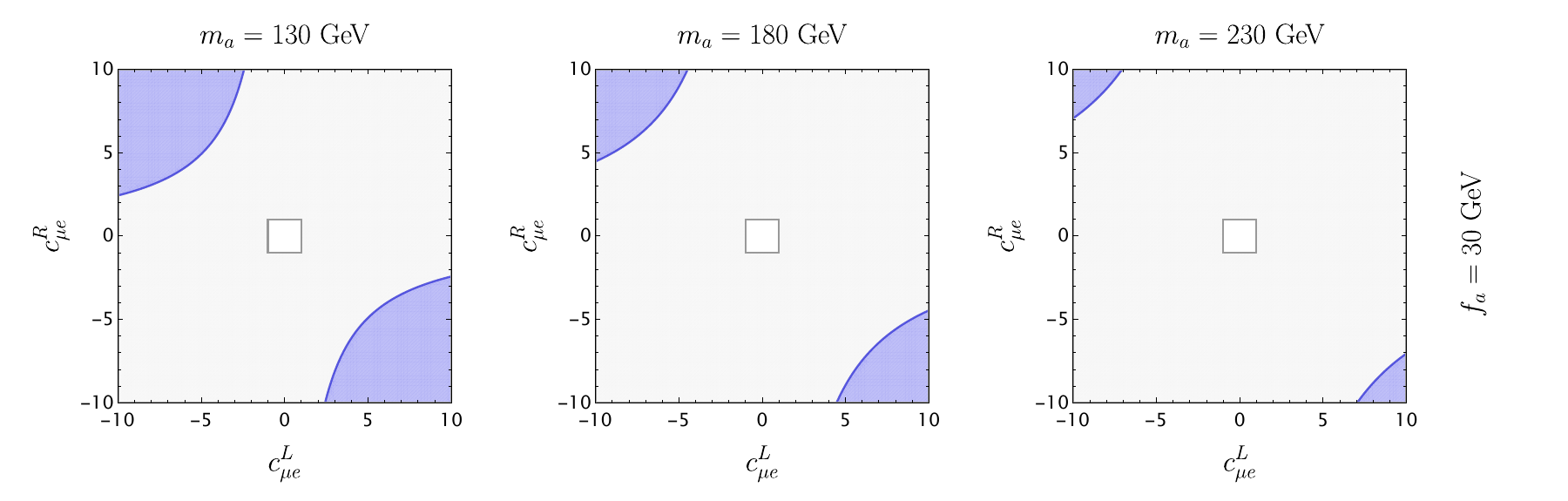}
\caption{\it Texture 2a). The $c^L_{\mu e}\times c^R_{\mu e}$ parameter space for fixed $c^{L,R}_{ii}=0$. $f_a=30\GeV$ in all the plots. $m_a=130(180)[230]\GeV$ in the plot on the left (centre) [right]. Blue regions are excluded at $3\sigma$ by the current experimental bounds. Gray region corresponds to $|c_{\mu e}^{L,R}|>1$.}
\label{clemucremutex2a}
\end{figure}

To better examine the low scales allowed by this texture, Fig.~\ref{clemucremutex2a} showcases three plots for the parameter space $c^{L,R}_{\mu e}$, assuming an extreme value for $f_a=30\GeV$ and progressively increasing values of $m_a$. The relevant aspect to be underlined refers to the values of the coefficients: the region for $\left|c^{L,R}_{\mu e}\right|\leq1$ is always unconstrained, for both present data and future prospects. Larger coefficient values, $c^{L,R}>1$,  would simply correspond to a redefinition of $f_a$ to even smaller values, and thus it would not represent a qualitatively different setup. We can therefore conclude that the suppression of the ALP-lepton couplings in this case, due to their proportionality to $m_e$ and $m_\mu$, leads to an undetectable ALP in any of the leptonic observables considered, even in the near future. 

\subsubsection*{Texture 2b)}

\begin{figure}[h!]
\includegraphics[width=1.05\textwidth]{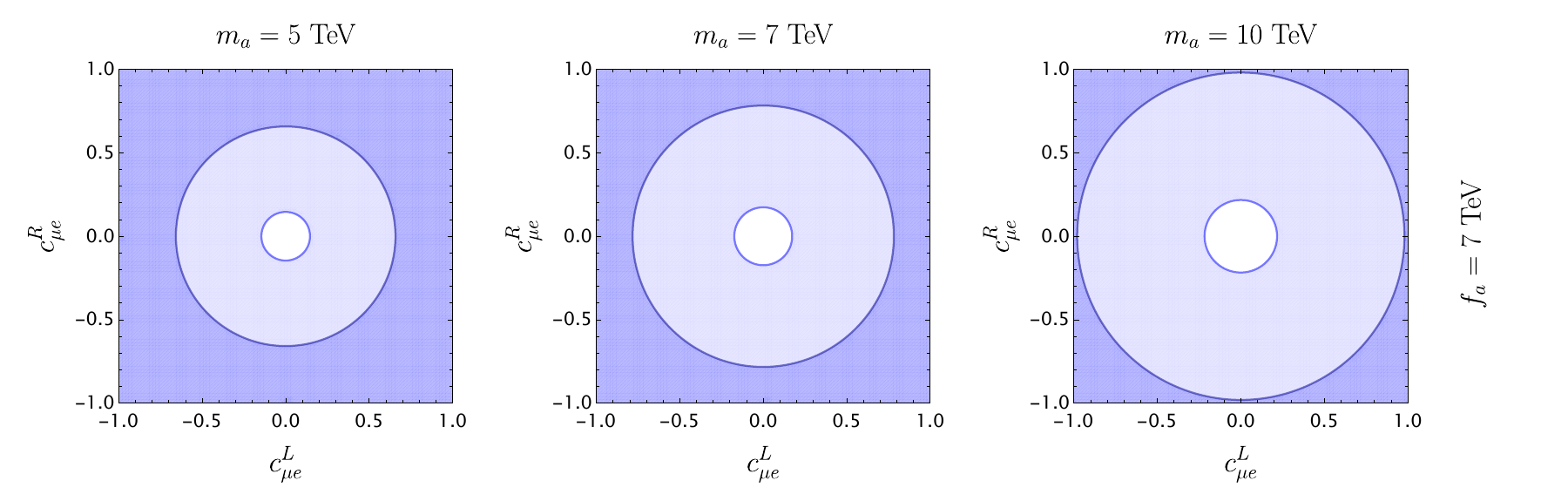}
\includegraphics[width=1.05\textwidth]{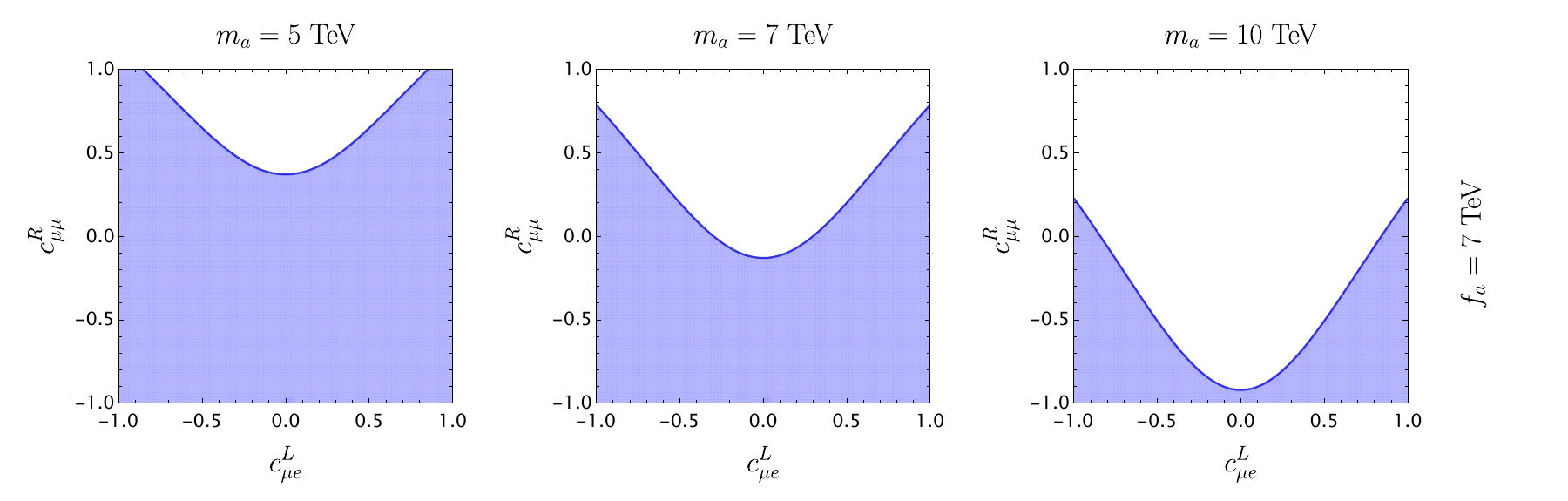}
\includegraphics[width=1.05\textwidth]{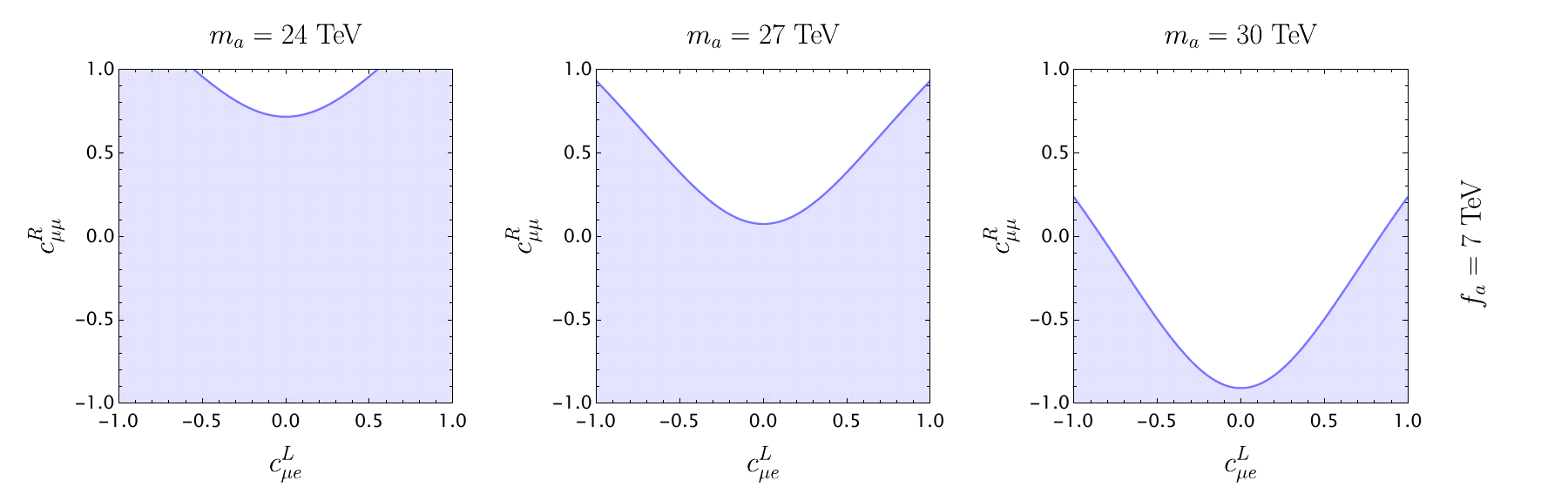}
\caption{\it Texture 2b). Top row: $c^L_{\mu e}\times c^R_{\mu e}$ parameter space for fixed $c^L_{ii}=1=-c^R_{ii}$. Central and lower rows: $c^L_{\mu e}\times c^R_{\mu\mu}$ parameter space for fixed $c^L_{ii}=1=-c^R_{ij}$. In all three cases, the chosen values of $m_a$ (on top of the plots) increase from left to right. $f_a=7\TeV$ in all the plots. The plots showcase the $3\sigma$ exclusion regions for the current (future) experimental bounds in darker (lighter) blue.}
\label{tex2bfp}
\end{figure}

This texture presents a democratic structure in the $e-\mu$ sectors, while the $\tau$-couplings are suppressed. The only relevant observables are the $g-2$ of electron and muon, and the decays of the muon. The most present experimental constraint is due to $\mu\to e\gamma$, while the dominant future prospect is from $\mNeN$. Indeed, this is reflected in the parameter space of the plots in Fig.~\ref{tex2bfp}. In order to understand the contours of the darker (lighter) blue regions we can focus on Eq.~\eqref{eq:br_mutoegamma} (Eq.~\eqref{eq:br_mue_conversion}). In particular, in the plots on the higher panel, we can identify the circular profile of the blue regions corresponding to the $\big(|c_{\mu e}^L|^2+|c_{\mu e}^R|^2\big)$ combination within the $\mathcal{BR}(\mu\to e\gamma)$ and $\mathcal{BR}(\mu^- N \to e^- N)$ expressions.

The profiles in the plots in the middle and lower panels can be explained by focussing on the flavour conserving couplings in Eqs.~\eqref{eq:br_mutoegamma} and \eqref{eq:br_mue_conversion}, namely the two relevant observables for the present constraints and future prospects, respectively. Recalling the definition of the effective photon coupling, the focus is on the combination $(c_{\mu\mu}^L - c_{\mu\mu}^R)^2$: fixing $c_{\mu\mu}^L=1$ implies the possibility of a partial cancellation of this combination upon variation of $c_{\mu\mu}^R$. The value of the observables Eqs.~\eqref{eq:br_mutoegamma} and \eqref{eq:br_mue_conversion} becomes suppressed, thus reducing the constraints on the corresponding parameter space. This behaviour is obtained when $c_{\mu\mu}^R\to 1$; however, when it turns negative, no cancellation is in action and the parameter space becomes excluded by both the present bounds and the future prospects.

The plots are presented for a specific value of $f_a$ chosen based on the plot in Fig.~\ref{Labelmafa12fpTexture2} in order to be close to the region where experimental bounds are relevant. Three values of $m_a$ are showcased in order to illustrate the variation of the parameter space for the given couplings as a function of $m_a$. While in the plots in the top row this dependence is trivial, in the others it is more involved: we observe how progressively increasing the masses (from left to right) opens up the available parameter space. The scaling of the bounds agrees with the expected behaviour dominated by the photonic penguin diagram contribution.

\subsubsection*{Texture 3a)}

This texture is a natural generalisation of the previous texture to the case of three generations and presents a completely democratic structure among its entries.
Since the $\tau$ contribution is switched on, $\tau$-decays can have a non-vanishing contribution, which is enhanced by $m_\tau/m_\mu$ with respect to the muon observables, although the former has lower precision than the latter. Performing our numerical analysis, the processes that yield the strongest present and future bounds are, respectively, $\mu\to e\gamma$ and $\mu N\to e N$.

\begin{figure}[h!]
\includegraphics[width=1.05\textwidth]{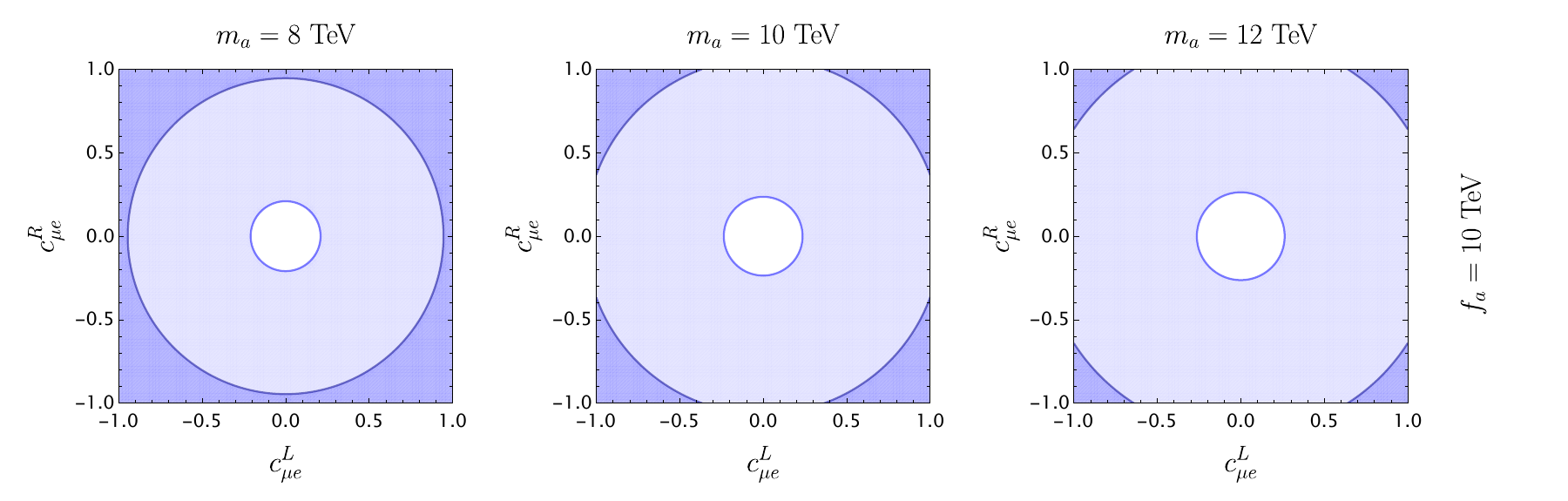}
\includegraphics[width=1.05\textwidth]{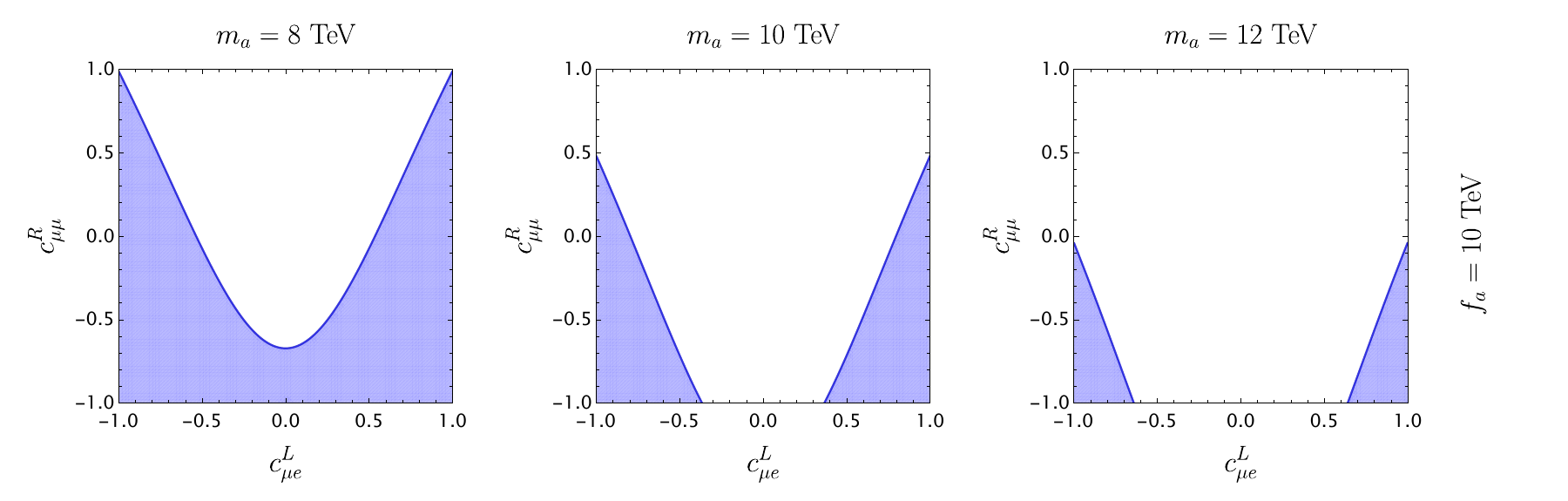}
\includegraphics[width=1.05\textwidth]{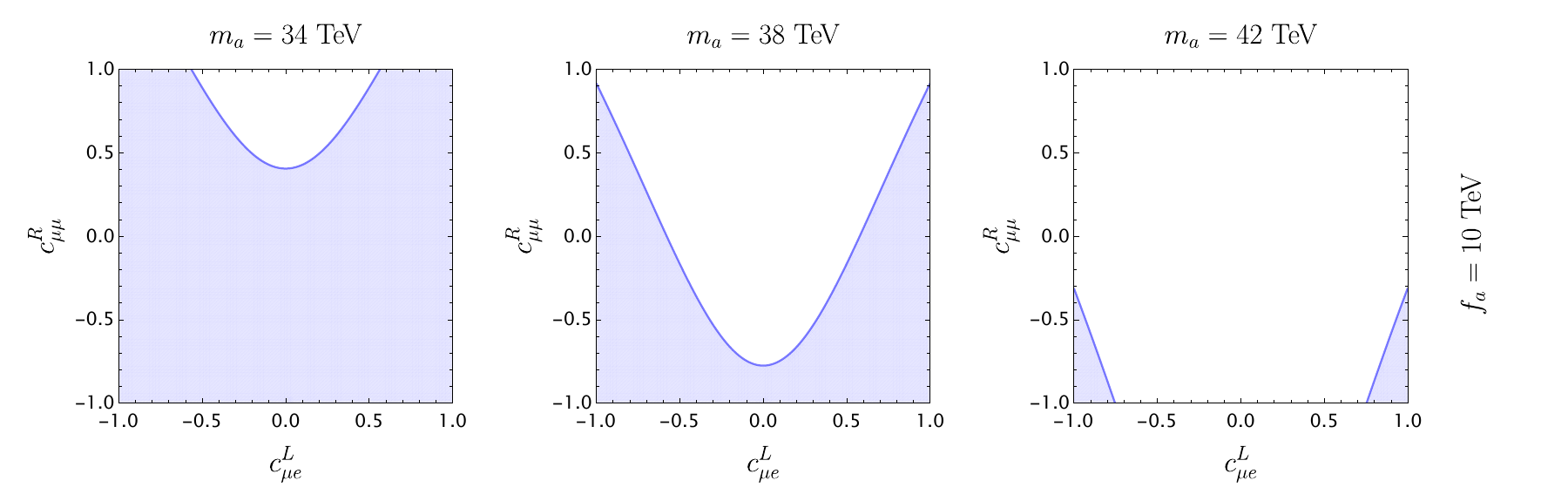}
\caption{\it Texture 3a). Top row: $c^L_{\mu e}\times c^R_{\mu e}$ parameter space for fixed $c^L_{ii}=1=-c^R_{ii}$. Central and lower rows: $c^L_{\mu e}\times c^R_{\mu\mu}$ parameter space for fixed $c^L_{ii}=1=-c^R_{ij}$. In all three cases, the chosen values of $m_a$ (on top of the plots) increase from left to right. $f_a=10\TeV$ in all the plots. The plots showcase the $3\sigma$ exclusion regions for the current (future) experimental bounds in darker (lighter) blue.} 
\label{tex3afp}
\end{figure}

Therefore, as in Texture 2b), the dominant contributions arise from the couplings $c_{\mu e}^{L,R}$ as well as the diagonal ones, due to the enhancement of the photon penguin terms. In Fig.~\ref{tex3afp} we provide the same plots as in Fig.~\ref{tex2bfp}, noticing the difference in the scale as a result of the contribution of $c_{\tau\tau}$ through $c_{a\gamma\gamma}^{\text{eff}}$ as already stated before.

\begin{figure}[h!]
\centering
\includegraphics[width=1.05\textwidth]{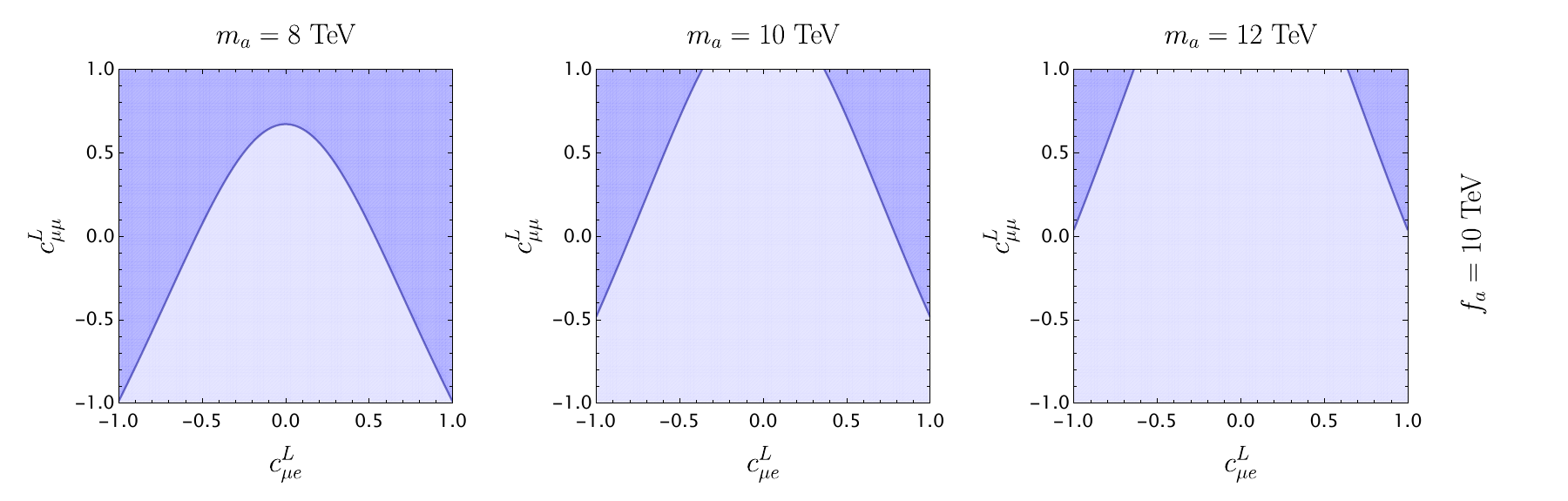}
\includegraphics[width=1.05\textwidth]{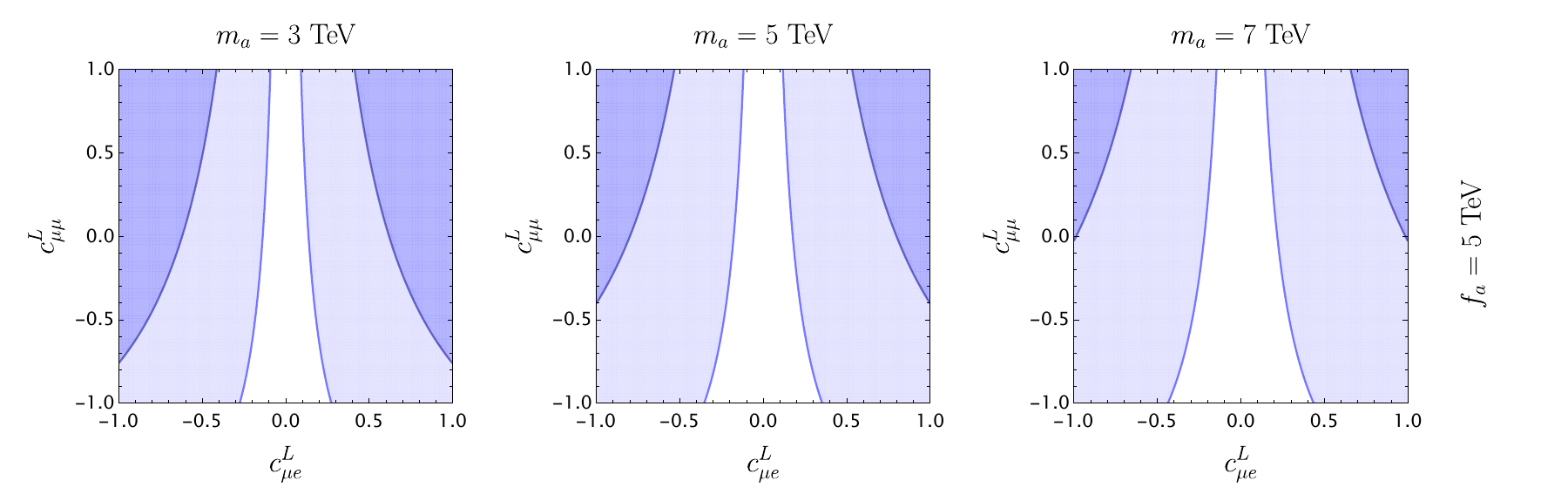}
\caption{\it Texture 3a). $c^L_{\mu e}\times c^L_{\mu \mu}$ parameter space (for fixed $c^L_{ii}=1=-c^R_{ii}$). Top row: $c^R=-1$ and $f_a = 10 \, \textrm{TeV}$. Bottom row: $c^R=0$ and $f_a = 5 \, \textrm{TeV}$. In all three cases, the chosen values of $m_a$ (on top of the plots) increase from left to right. The plots showcase the $3\sigma$ exclusion regions for the current (future) experimental bounds in darker (lighter) blue.}
\label{Labelcletaucltaumutex1chiral}
\end{figure}

For this texture, we also consider the case where all the couplings of one chirality vanish. In this case, the only surviving contributions are those proportional to products of couplings with the same chirality, such as those in the second line of Eq.~\eqref{eq:muegammatau} for $\mathcal{BR}(\mu\to e\gamma)$. The plots in Fig.~\ref{Labelcletaucltaumutex1chiral} explicitly show how the parameter space varies switching off the $c^R$ coupling: in the higher panel $c^R=-1$ while in the lower one $c^R=0$. The complementary case with $c^L=0$ is equivalent. The scales involved in the plots in Fig.~\ref{Labelcletaucltaumutex1chiral} are smaller than those in Fig.~\ref{tex3afp} (which are equivalent plots due to the symmetry between $c^L$ and $c^R$) as the dominant contributions of the relevant processes halved and, therefore, the experimental bounds become weaker.

\subsubsection*{Texture 3b)}

Adopting the same coupling structure as in the previous Texture 3a) a suppression of $\mathcal{O}(10^{-3})$ in the tau sector effectively reduces the theory to a two-lepton scenario where only ALP couplings with electrons and muons are relevant. The processes that provide the strongest constraints on the parameter space remain the same as in the previous Texture 2b) and 3a) and the plots in Fig.~\ref{tex2bfp} also apply to the Texture 3b) case.

The contribution from diagonal ALP-$\tau$ couplings represents no major impact on $c_{a\gamma\gamma}^{\text{eff}}$ making this texture essentially equivalent to Texture 2b). On that line, it is useful to comment that if the $\tau$-entries were \mbox{$\sim\mathcal{O}(10^{-1})$}, the contribution on  $c_{a\gamma\gamma}^{\text{eff}}$ would be able to represent an increment in the scale with respect to the Texture 2b). On the contrary, if these entries were an order of magnitude smaller (\mbox{$\sim\mathcal{O}(10^{-4})$}), the corresponding texture would receive contributions from the RGE running, effectively leading at low-energy to the original texture 3b).

\subsubsection*{Texture 3c)}

\begin{figure}[h!]
\includegraphics[width=1.05\textwidth]{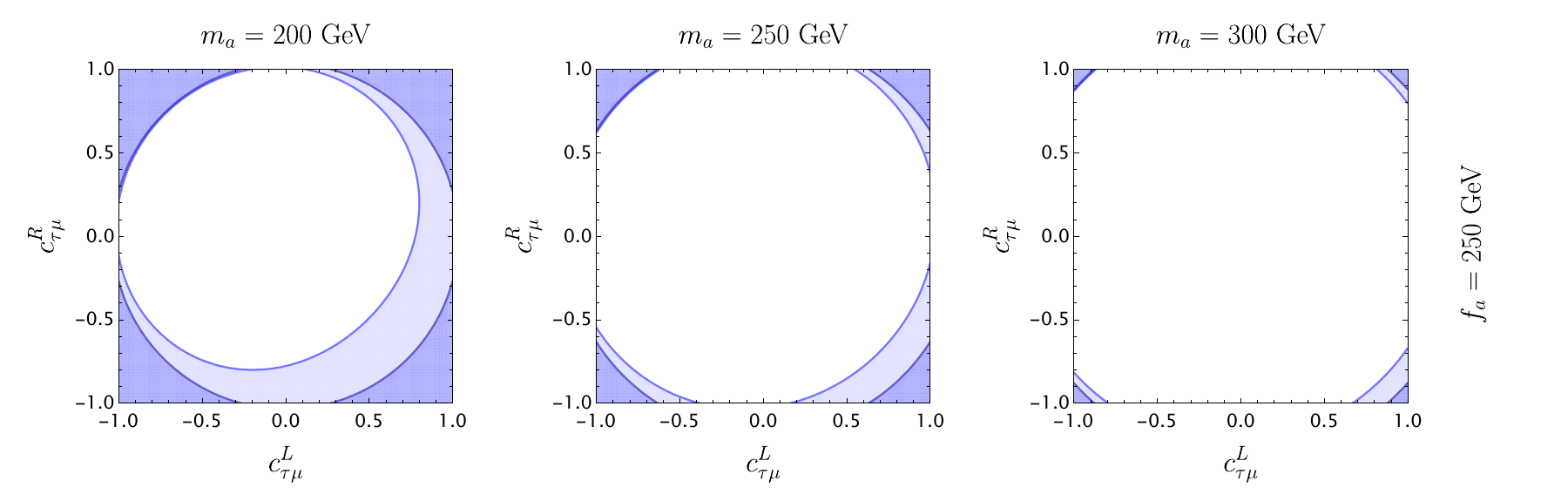}
\includegraphics[width=1.05\textwidth]{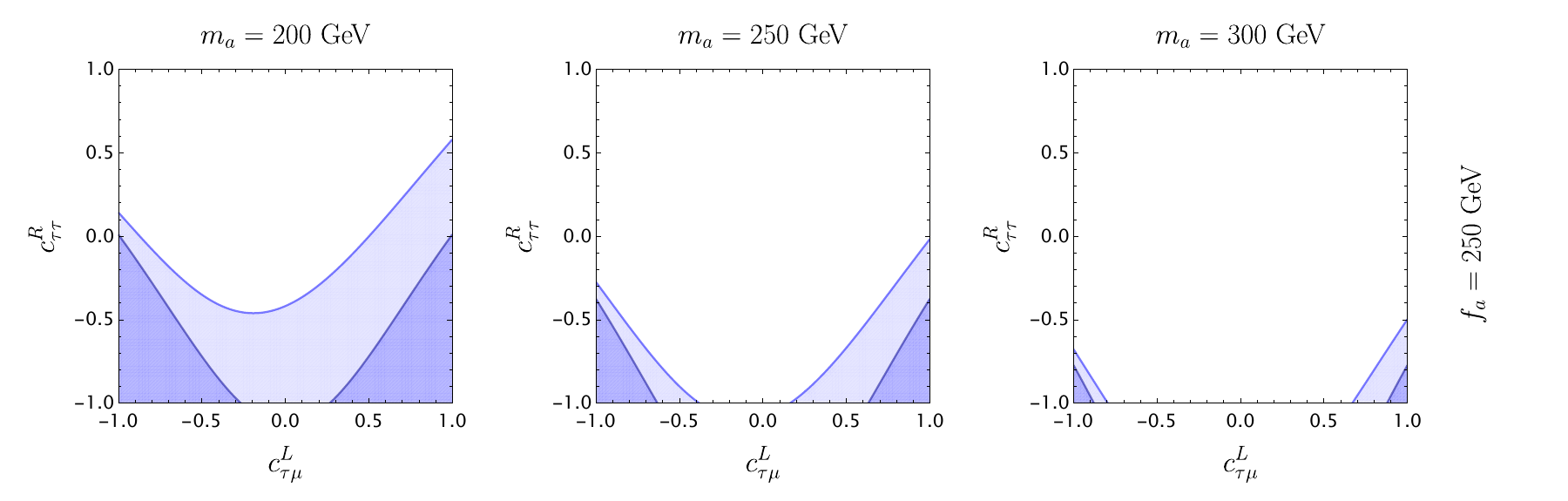}
\caption{\it Texture 3c). Top row: $c^L_{\tau \mu}\times c^R_{\tau \mu}$ parameter space. Bottom row: $c^L_{\tau \mu}\times c^R_{\tau \tau}$ parameter space. While the couplings that appear in the axes that are kept free, those ones that refer to the electron are fixed to $c^L=10^{-4}=-c^R$ and the rest are fixed to $c^L=1=-c^R$. $f_a=250\GeV$ in all the plots. $m_a=200(250)[300]\GeV$ in the left (centre) [right] column. The plots showcase the $3\sigma$ exclusion regions for the current (future) experimental bounds in darker (lighter) blue.}
\label{tex3cfp}
\end{figure}

This analysis also focusses on a two-family framework, suppressing all terms related to the electron by four orders of magnitude compared to the other terms. This choice is made intentionally, as the process $\meg$ is no longer the most constraining, having been surpassed by $\tmg$ for present bounds.
For future prospects, the process $\mNeN$ improves significantly with respect to the current situation, but still cannot provide the strongest bound, although it competes with $\tmg$.

The dominant contributions arise from the couplings $c_{\tau\mu}^{L,R}$ and $c_{\tau\tau}$ due to the enhancement of the photon penguin terms. However, due to the suppression of the muon observables and the lesser precision in the $\tau$ ones, the scale is expected to be lower than the analogue Texture 3b).

In the plots in Fig.~\ref{tex3cfp} we illustrate the parameter space $c^L_{\tau\mu}\times c^R_{\tau\mu}$ on the higher panel, and $c^L_{\tau\mu}\times c^R_{\tau\tau}$ on the lower one.
The shape of the different plots is equivalent to the ones provided in Fig.~\ref{tex2bfp}, only changing the scale, as stated previously. Since the future prospects of $\mNeN$ can at best compete with $\tmg$, the change in shape and scale in both plots is minimal.

The ALP-$e$ couplings have negligible effects in the $\chi^2$ and as a result no meaningful conclusion can be deduced from plots focussing on the $c^L_{ie}\times c^R_{ie}$ parameter space.

As commented for the previous texture, smaller values of the flavour-violating entries would lead to quantum corrections running from the high-energy scale down to $m_a$, while texture 3c) is stable under RGE running. Increasing the suppressed entries by an order of magnitude would recover the processes $\mu\to e\gamma$ and $\mNeN$ as the more constraining ones for present bounds and future sensitivities, respectively.

\subsubsection*{Texture 3d)}

In this texture, all the couplings associated with the muon are $\cO(10^{-4})$, representing the complementary framework with respect to textures 3b) and 3c). 

\begin{figure}[h!]
\includegraphics[width=1.05\textwidth]{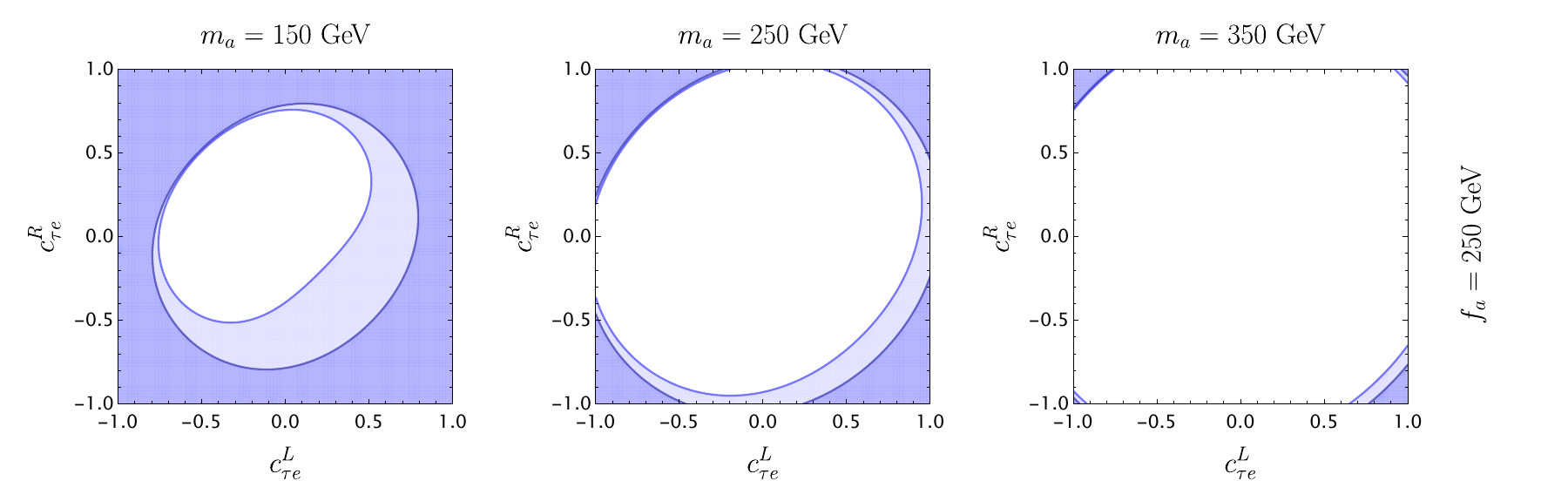}
\includegraphics[width=1.05\textwidth]{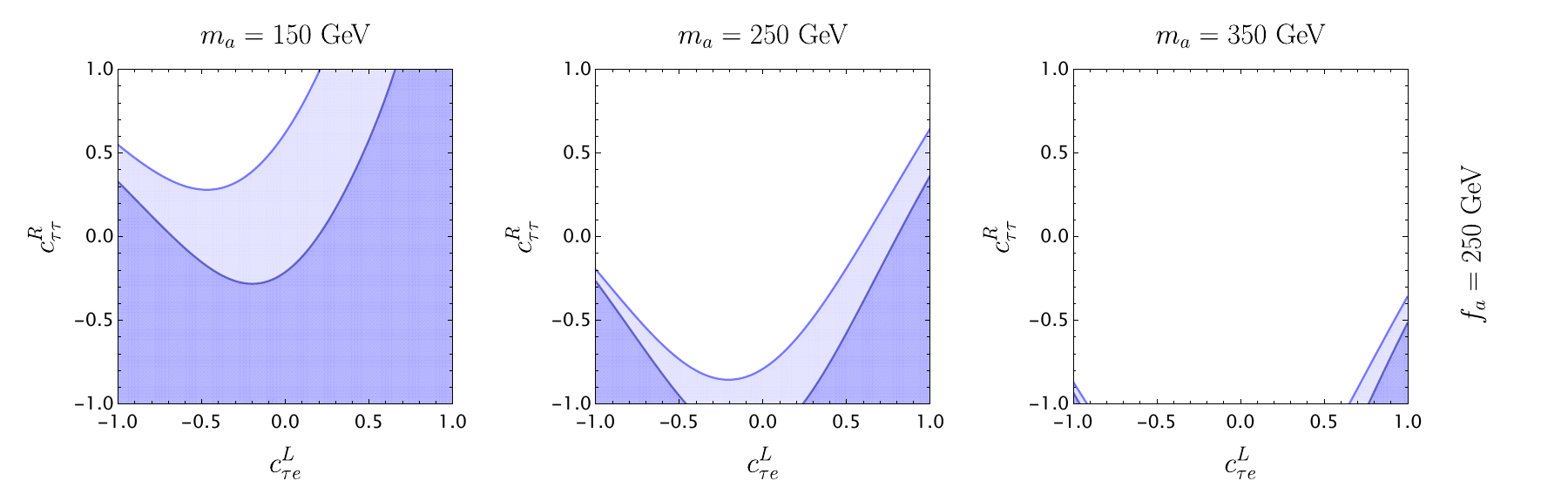}
\caption{\it Texture 3d). Top row: $c^L_{\tau e}\times c^R_{\tau e}$ parameter space. Bottom row: $c^L_{\tau e}\times c^R_{\tau \tau}$ parameter space. While the couplings that appear in the axes that are kept free, those ones that refer to the muon are fixed to $c^L=10^{-4}=-c^R$ and the rest are fixed to $c^L=1=-c^R$. $f_a=250\GeV$ in all the plots. $m_a=200(250)[300]\GeV$ in the left (centre) [right] column. The plots showcase the $3\sigma$ exclusion regions for the current (future) experimental bounds in darker (lighter) blue.}
\label{tex3dfp}
\end{figure}

The discussion of this texture becomes equivalent to the previous one, but replacing $\tmg$ by $\teg$ as the process that yields the strongest constraints. For future prospects, the process $\mNeN$ can compete with $\teg$ but becomes less stringent than for $\tmg$ in the previous texture.

The dominant contributions arise from the couplings $c_{\tau e}^{L,R}$ and $c_{\tau\tau}$, but are not necessary from the photon penguin terms. In fact, a remarkable contribution for this texture arises from the $(g-2)_e$ observable, as the scales involved make the photonic contribution no longer relevant compared with the non-photonic ones. That contribution, being $\Re[c_{\tau e}^L c_{\tau e}^R]$ in Eq.~\eqref{eq:g2electrontau}, is responsible for altering the shape of the plots in Fig.~\ref{tex3dfp} compared to the ones in Fig.~\ref{tex3cfp}. For the future prospects, the lesser constraints of $\mNeN$ can slightly change the shape and scale in both plots.

The ALP-$\mu$ couplings have negligible effects in the $\chi^2$ and therefore plots illustrating the parameter space $c^L_{\mu i}\times c^R_{\mu i}$ would not provide insightful results. 

This texture is also stable under RGE running effects, while smaller entries than $\sim10^{-4}$ would be affected by quantum corrections. As for the Texture 3c), increasing the suppressed entries by an order of magnitude would recover the processes $\mu\to e\gamma$ and $\mNeN$ as the more constraining ones for present bounds and future sensitivities, respectively.

\boldmath
\subsection{The $a^\text{Cs}_e$ Anomaly}
\unboldmath

In the previous analysis we considered the Rubidium determination for $(g-2)_e$ that presents a tension with the present experimental measurement of $+2.4\sigma$. This is a more conservative approach, while if we adopt the Caesium determination the tension increases up to $-3.8\sigma$ and thus represents an anomaly. In this section, we discuss whether and under which conditions a leptophilic ALP can explain this tension. As already mentioned in our analysis, which rests on the consistency requirement that $f_a>v_\text{EW}/4\pi$, we particularly focus on the $m_a>f_a$ parameter space. This is complementary to other studies that instead focus on the opposite regime, see for example, Refs.~\cite{Bauer:2019gfk,Cornella:2019uxs}.

The expression for the dominant ALP contribution to the $(g-2)_e$ is Eq.~\eqref{eq:ae} in the $e-\mu$ family case and Eq.~\eqref{eq:g2electrontau} for the three-family framework. Both flavour violating and conserving ALP-lepton couplings enter these expressions. The first ones have a chiral enhancement that is $m_\mu/m_e$ and $m_\tau/m_e$, respectively, due to the pure lepton-mediated diagram. This does not occur for the flavour conserving ones, associated to the photon penguin contributions. On the other hand, the latter have a milder $m_a$ dependence and turn out to be dominant at large ALP masses.

\begin{figure}[h!]
\centering
\includegraphics[width=0.45\textwidth]{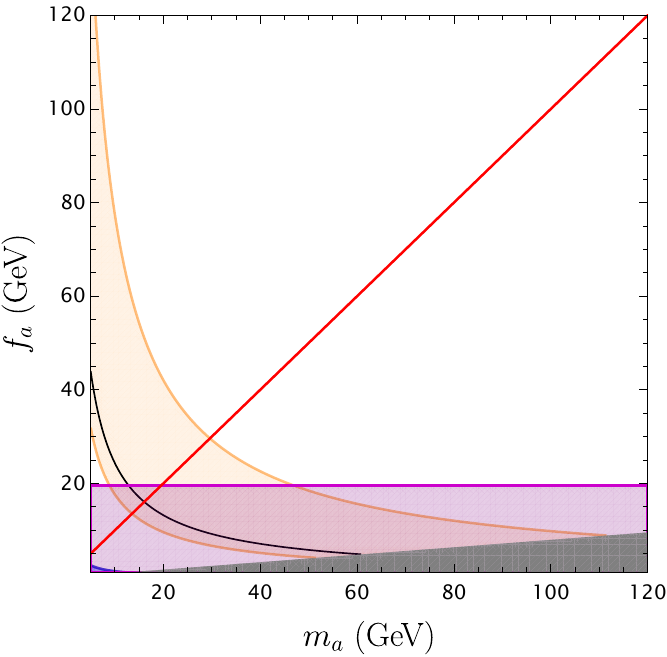}
\includegraphics[width=0.467\textwidth]{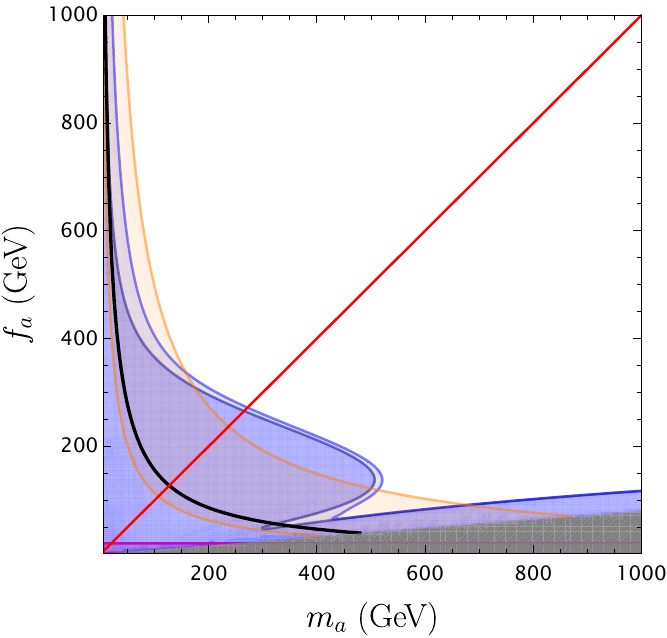}
\caption{\it $m_a\times f_a$ parameter space. The orange region identifies the parameter space where the Caesium anomaly can be solved at $3\sigma$, with the black line representing the best fit values. The left plot refers to the texture 2a), while the right plot to the texture 3d). The darker (lighter) blue regions are excluded by the present experimental bound (by the future prospects) at $3\sigma$.  The red line represents the condition $m_a=f_a$. The gray and purple regions are excluded by the theoretical consistency conditions $m_a<4\pi f_a$ and $f_a< v_{\text{EW}}/4\pi$, respectively.}
\label{CaesiumAnomaly}
\end{figure}

The plots in Fig.~\ref{CaesiumAnomaly} show our results for Texture 2a) within the $e$–$\mu$ framework (left) and for Texture 3d) within the three-family setup (right). In all panels, the orange region denotes the values of the parameter space $m_a \times f_a$ that resolve the Caesium tension at $3\sigma$, assuming $c^L_{\mu e}=1=-c^R_{\mu e}$ in the $e$–$\mu$ plot and $c^L_{\tau e}=1=-c^R_{\tau e}$ in the three-family one. Present experimental bounds and future sensitivities are shown in blue -- notice the barely visible blue region on the lower left corner of the plot on the left.

Starting with the $e$–$\mu$ scenario, an ALP obeying Texture 2a) can indeed account for the Caesium anomaly while remaining fully compatible with current and projected constraints. The only additional requirement to keep in mind is $f_a > v_\text{EW}/4\pi$, which implies that only the region with $m_a < f_a$ is consistent with our effective description. The analogous analysis for Texture 2b) excludes an ALP explanation of the anomaly, as the $\meg$ limits impose a much stronger lower bound on both $m_a$ and $f_a$.

Turning now to the three-family case with Texture 3d), we find that present data, dominated by $\teg$, exclude this solution over a large portion of the parameter space, though not entirely. An ALP with $m_a \approx 500\GeV$ and $f_a \approx 100\GeV$, or alternatively with $m_a \approx 100\GeV$ and $f_a \gtrsim 400\GeV$, evades current $\teg$ bounds and future $\mNeN$ searches while reducing the $(g-2)_e$ tension to the $1\sigma$ level. In the region $m_a>f_a$, the allowed parameter space arises from the cancellation between the pure lepton mediated contribution and the photon penguin one.

Considering now the other three-generation textures, this ALP solution becomes strongly excluded. Indeed, once the $e$–$\mu$ couplings are switched on, then the $\meg$ and $\mNeN$ bounds are much more relevant, thus excluding the region of the parameter space $m_a\times f_a$ interesting for this anomaly.

\section{Final Remarks}
\label{sec:Conclusions}

In this work we have revisited one of the most widely used assumptions in the ALP literature, namely the parametric hierarchy $m_a \ll f_a$. While this relation is often taken as a defining feature of ALP effective descriptions, it is not a fundamental consistency requirement of the EFT framework. The meaningful theoretical constraint is instead $m_a < \Lambda$, where $\Lambda$ denotes the true EFT cut-off. The comparison between $m_a$ and $f_a$ is therefore model-dependent and, in general, does not carry intrinsic physical significance. Moreover, using NDA, the region $m_a > f_a$ is naturally interpreted as pointing towards a strongly coupled ultraviolet completion.

Motivated by this theoretical reassessment, we have performed a dedicated phenomenological analysis of leptophilic ALPs in the regime $m_a > f_a$, a region that has received little attention in previous studies. Additionally, we have implemented the full RGE from the cut-off scale down to $m_a$. Integrating out the considered leptophilic ALP and matching onto the SMEFT Warsaw basis, we could conclude that all induced quark operators are strongly suppressed, both by Yukawa insertions and by loop effects. This establishes that the hadronic sector does not play a significant role in the phenomenology under consideration, thereby justifying our focus on purely leptonic observables. On the other hand, we have re-derived the one-loop expressions for the anomalous magnetic dipole moments of the electron and muon, as well as for radiative and three-body charged lepton flavour-violating decays and $\mu$--$e$ conversion in nuclei, for both the two-family and three-family frameworks. In the latter case, we have shown explicitly how $\tau$-induced contributions can dominate due to the $m_\tau$ enhancement, even when the corresponding flavour couplings are comparatively small. The interplay between tree-level, one-loop and interference terms has been carefully assessed for large ALP mass, resulting in a study complementary to existing literature.  

From the phenomenological perspective, our analysis demonstrates that sizeable regions of parameter space with $m_a > f_a$ remain viable under current experimental constraints. Moreover, this region can lead to distinctive patterns in charged lepton flavour violation and in the anomalous magnetic moments. In particular, if no specific hierarchy is considered in the ALP-lepton couplings and they are taken $\cO(1)$, then the present experimental constraints, dominated in general by the $\meg$ process, imply the condition $m_a f_a\gtrsim(12\TeV)^2$. Otherwise, if the flavour violating couplings in any of the flavour sectors are suppressed by one order of magnitude, this bound gets very much relaxed, $m_a f_a\gtrsim(4\TeV)^2$. Considering future prospects, dominated by the $\mNeN$ conversion, these conditions are slightly stronger, $m_a f_a\gtrsim(25\TeV)^2$ and $m_a f_a\gtrsim(8\TeV)^2$, respectively. We have also considered specific benchmark scenarios for the ALP-lepton couplings to investigate the dependence of the parameter space on the flavour violating ALP couplings. Specifically, we focussed on electron-phobic, muon-phobic, and tau-phobic frameworks, assessing the stability of the assumed textures under RGE running and clearly concluding that the region of parameter space with $m_a>f_a$ is very much unconstrained under both present experimental data and future prospects.

Furthermore, we have identified the conditions under which the existence of an ALP can accommodate the Caesium determination of $(g-2)_e$, which currently presents a $-3.8\sigma$ tension with the corresponding experimental measure. In the two-family case of a tau-phobic ALP, we found that only by suppressing the flavour conserving ALP-coupling to the muon can the ALP hypothesis explain the present tension, as in this way it is possible to evade the $\meg$ and $\mNeN$ limits. The values of the ALP mass and decay constant $f_a$ have to satisfy the relation $(10\GeV)^2\lesssim m_a f_a\lesssim(30\GeV)^2$, together with the consistency condition $f_a>v_\text{EW}/4\pi$. In the three-family scenario, we highlighted the fundamental role played by the flavour violating couplings $c^{L,R}_{\tau e}$, and by the relative chiral structure of the ALP-lepton interactions with their proportionality to the lepton masses. We concluded that a leptophilic ALP can indeed solve the anomaly in the Caesium determination of $(g-2)_e$ if its couplings to muons are suppressed with respect to those with taus and electrons: we identified two kind of possibilities, the first being an ALP with $m_a\approx500\GeV$ and $f_a\approx100\GeV$, and the second  for $m_a\approx100\GeV$ and $f_a\gtrsim400\GeV$. 

Overall, our results show that the region $m_a > f_a$ is not only theoretically consistent but also phenomenologically rich and experimentally testable. The widespread identification of $f_a$ with the EFT cut-off has led to an overly restrictive exploration of ALP parameter space. Relaxing this prejudice opens new avenues both for model building and for experimental searches, particularly in precision leptonic observables.

Abandoning the leptophilic scenario considered here, one has to take into consideration colliders bounds associated to the ALP-gauge field couplings and ALP-top interactions, and the various constraints arising from the hadronic sector. These studies are complementary with respect to our analysis and are deferred to a future investigation.

\section*{Acknowledgements}
We are grateful to Emmanuel Stamou, Enrique Fern\'andez Mart\'inez and Pablo Qu\'ilez for valuable discussions.

The work of MFZ is supported by the Spanish MIU through the National Program FPU (grant number FPU22/\hspace{0pt}03625). The work of ALO is supported by the FPI grant PRE2022-105383 funded by MCIN/AEI/10.13039/501100011033 and ESF+. The work of SRH is supported by the FPI grant PREP2022-000805 funded by MCIN/AEI/10.13039/501100011033.

We acknowledge partial financial support by the European Union's Horizon 2020 research and innovation programme under the Marie Sk\l odowska-Curie grant agreement No.~101086085-ASYMMETRY and by the Spanish Research Agency (Agencia Estatal de Investigaci\'on) through the grant IFT Centro de Excelencia Severo Ochoa No CEX2020-001007-S and by the grant PID2022-137127NB-I00 funded by MCIN/\hspace{0pt}AEI/\hspace{0pt} 10.13039/\hspace{0pt}501100011033.
This article is based upon work from COST Action COSMIC WISPers CA21106, supported by COST (European Cooperation in Science and Technology).\\

\appendix

\section{Running and Matching of the ALP Couplings}
\label{app:Running}
In Sect.~\ref{sec:flavour_textures}, we discuss the different coupling textures (see \Cref{eq:texture_2a,eq:texture_2b,eq:texture_3a,eq:texture_3b,eq:texture_3c,eq:texture_3d}) used in our phenomenological analysis. A question one might have is whether the running would noticeably impact the couplings, thus completely  altering the coupling texture at low energies, at which the process is computed at which the processes of our analysis are computed and compared with experiment. Here we will show that this is not the case. For this, we run the couplings from a set scale $\Lambda=10\TeV$, corresponding to the highest scale present in our analysis, down to the mass of the ALP, $m_a\in \left[50,400\right]\GeV$ using \alpaca~\cite{Alda:2025nsz}. For energies scales below the ALP mass, where the ALP is already integrated out, we do not account for the EW running, since its contributions are negligible. In the following, we report the order of magnitude of the couplings associated with the different textures. As initial values, we assume that the diagonal elements of $c^L$ and $c^R$ have opposite signs, while for the off-diagonal elements we assume them to be equal.
\paragraph{Texture 2a:}
\begin{equation}
\label{eq:texture_2a_run}
\qquad\qquad    
c^{L/R}= \left( \begin{array}{cc}
        0 &1\\
        1 &  0\\
    \end{array}\right)\xrightarrow{\text{RGE}}|c^{L/R}|\sim\left( \begin{array}{cc}
        0 &\mathcal{O}(1)\\
        \mathcal{O}(1) &  0\\
    \end{array}\right)\,
\end{equation}
\paragraph{Texture 2b):}
\begin{equation}
\label{eq:texture_2b_run}
\qquad\qquad   
c^{L/R}= \left( \begin{array}{cc}
        \pm 1 &1\\
        1 & \pm 1\\
    \end{array}\right)\xrightarrow{\text{RGE}}|c^{L/R}|\sim \left( \begin{array}{ccc}
        \mathcal{O}(1) &\mathcal{O}(1)&0\\
        \mathcal{O}(1) & \mathcal{O}(1)&0\\
        0&0&\mathcal{O}(10^{-4})\\
    \end{array}\right)\,
 \end{equation}
\paragraph{Texture 3a):}
\begin{equation}
\qquad\qquad  
    c^{L/R}= \left( \begin{array}{ccc}
        \pm 1 &1 &1 \\
        1 & \pm 1&1\\
        1 & 1&\pm 1
    \end{array}\right)\xrightarrow{\text{RGE}}|c^{L/R}|\sim \left( \begin{array}{ccc}
        \mathcal{O}(1) &\mathcal{O}(1) &\mathcal{O}(1) \\
        \mathcal{O}(1) & \mathcal{O}(1)&\mathcal{O}(1)\\
        \mathcal{O}(1) & \mathcal{O}(1)&\mathcal{O}(1)
    \end{array}\right)
\end{equation}
\paragraph{Texture 3b):}
\begin{flalign}
\hspace{-1.75cm}
\qquad\qquad  
    c^{L/R}=\left( \begin{array}{ccc}
        \pm 1 &
        1& 
        10^{-3} \\
        1 & 
        \pm 1&
        10^{-3}\\
        10^{-3}&
        10^{-3}&
        \pm 10^{-3}
    \end{array}\right)\xrightarrow{\text{RGE}} |c^{L/R}|\sim\left( \begin{array}{ccc}
        \mathcal{O}(1) &
        \mathcal{O}(1)& 
        \mathcal{O}(10^{-3}) \\
        \mathcal{O}(1) & 
        \mathcal{O}(1)&
        \mathcal{O}(10^{-3})\\
        \mathcal{O}(10^{-3})&
        \mathcal{O}(10^{-3})&
        \mathcal{O}(10^{-3})
    \end{array}\right)
\end{flalign}
\paragraph{Texture 3c):}
\begin{equation}
\hspace{-1.75cm}
\qquad\qquad  
     c^{L/R}= \left( \begin{array}{ccc}
        \pm 10^{-4} &10^{-4}&10^{-4} \\
        10^{-4} & \pm 1&  1\\
        10^{-4}&  1& \pm 1
\end{array}\right)\xrightarrow{\text{RGE}} |c^{L/R}|\sim \left( \begin{array}{ccc}
        \mathcal{O}(10^{-4}) &
        \mathcal{O}(10^{-4})&
        \mathcal{O}(10^{-4}) \\
        \mathcal{O}(10^{-4}) & 
        \mathcal{O}(1)&
        \mathcal{O}(1)\\
        \mathcal{O}(10^{-4})&
        \mathcal{O}(1)&
        \mathcal{O}(1)
    \end{array}\right)\,
\end{equation}
\paragraph{Texture 3d):}
\begin{equation}
\hspace{-1.75cm}
\qquad\qquad 
     c^{L/R}\sim \left( \begin{array}{ccc}
        \pm 1 &
        10^{-4}&
        1 \\
        10^{-4} & 
        \pm 10^{-4}&
        10^{-4}\\
        1&
        10^{-4}&
        \pm 1
    \end{array}\right)\xrightarrow{\text{RGE}}|c^{L/R}|\sim \left( \begin{array}{ccc}
        \mathcal{O}(1) &
        \mathcal{O}(10^{-4})&
        \mathcal{O}(1) \\
        \mathcal{O}(10^{-4}) & 
        \mathcal{O}(10^{-4})&
        \mathcal{O}(10^{-4})\\
        \mathcal{O}(1)&
        \mathcal{O}(10^{-4})&
        \mathcal{O}(1)
    \end{array}\right)\,
\end{equation}
Notice that after RGE the couplings alter their exact values but not the overall order of magnitude. Additionally, in the derivative basis, the ALP coupling to gauge bosons is scale-invariant. Therefore, they will not be generated via running.

Another relevant point we briefly discussed was the fact that we do not need to consider observables with quarks since the Wilson coefficients associated to the integration out of the ALP, Eq.~\ref{eq:wilson_coefficient}, are all suppressed with the exception of $\mathcal{C}_{le}$. Such smallness is related to the fact that the quark $\widehat{Y}^\prime$ are suppressed by the Yukawa and the ALP--quark couplings generated via running. Indeed, an explicit calculation shows that the biggest contributions to the quark $\widehat{Y}^\prime$ are related to the $t$ and $b$ quarks, as could be expected. The dominant contribution to this quantity for the different textures can be seen in Tab.~\ref{tab:Running2Families} in the two-family case, and Tab.~\ref{tab:Running3Families} for the three-family case. For these calculations, we fix the input parameters $\Lambda=10\TeV$ and $m_a\in \left[100,400\right]\GeV$.

In Eq.\ (\ref{eq:wilson_coefficient}) we demonstrate that these quantities enter the Wilson coefficients in pairs and are additionally suppressed by $v^2/(m_a^2 f_a^2)$. Therefore, all operators involving quarks will be suppressed at least by a factor ~$10^{-6}v^2/(m_a^2 f_a^2)$.

\begin{table}[h!]
    \centering
    \begin{tabular}{c|c}
   & Dominant $\widehat{Y}'(m_a)$\\[1mm]
   \hline\\
   &\\[-6mm]
         Texture 2a) & $(\widehat{Y}'_d)=0,\,(\widehat{Y}'_u)=0,\,(\widehat{Y}'_e)^{12}\sim (\widehat{Y}'_e)^{21}\sim 10^{-4}$\\[1mm]
         Texture 2b) &  $(\widehat{Y}'_d)^{33}\sim 10^{-6},\,(\widehat{Y}'_u)^{33}\sim 10^{-4},\,(\widehat{Y}'_e)^{22}\sim 10^{-3}$\\
    \end{tabular}
    \caption{\it $e-\mu$ framework. Dominant contribution to $\widehat{Y}'(m_a)$, performing the RGE running from $\Lambda=10\TeV$ down to $m_a\in \left[100,400\right]\GeV$, using \alpaca. The initial values of the couplings assume the LH and RH couplings to have opposite (equal) signs (positive for LH, negative for RH) for the diagonal (off-diagonal) terms.}
    \label{tab:Running2Families}
\end{table}

\begin{table}[h!]
    \centering
    \begin{tabular}{c|c}
   & Dominant $\widehat{Y}'(m_a)$\\[1mm]
   \hline\\
   &\\[-6mm]
         Texture 3a)&  $(\widehat{Y}'_d)^{33}\sim 10^{-5},\,(\widehat{Y}'_u)^{33}\sim 10^{-4},\,(\widehat{Y}'_e)^{33}\sim 10^{-2}$\\[1mm]
         Texture 3b)&  $(\widehat{Y}'_d)^{33}\sim 10^{-6},\,(\widehat{Y}'_u)^{33}\sim 10^{-4},\,(\widehat{Y}'_e)^{22}\sim 10^{-3}$\\[1mm]
         Texture 3c)&   $(\widehat{Y}'_d)^{33}\sim 10^{-6},\,(\widehat{Y}'_u)^{33}\sim 10^{-4},\,(\widehat{Y}'_e)^{33}\sim 10^{-2}$\\[1mm]
         Texture 3d)&   $(\widehat{Y}'_d)^{33}\sim 10^{-6},\,(\widehat{Y}'_u)^{33}\sim 10^{-4},\,(\widehat{Y}'_e)^{33}\sim 10^{-2}$
    \end{tabular}
    \caption{\it Three-family framework. 
    Dominant contributions to $\widehat{Y}'(m_a)$, performing the RGE running from $\Lambda=10\TeV$ down to $m_a\in \left[100,400\right]\GeV$, using \alpaca. The initial values of the couplings assume the LH and RH couplings to have opposite (equal) signs (positive for LH, negative for RH) for the diagonal (off-diagonal) terms.}
    \label{tab:Running3Families}
\end{table}


\bibliographystyle{utphys}
\bibliography{references}

\end{document}